\documentclass[12pt]{article}
\usepackage[textwidth=6.25in]{geometry}

\usepackage{enumerate}
\usepackage{amsmath} 
\usepackage{times}
\usepackage{amsmath,amsthm,amssymb}
\usepackage{fancyhdr}
\usepackage{moreverb}
\usepackage{graphicx}
\usepackage{amssymb}
\usepackage{url}
\usepackage{multirow} 
\usepackage[boxed]{algorithm}
\usepackage{algpseudocode}
\usepackage{multirow,bigdelim}
\usepackage{geometry}
\usepackage{fix-cm}
 \usepackage{setspace}
\usepackage{natbib}
\usepackage{color}
\usepackage{bbold}
\usepackage{graphicx}
\usepackage{caption}
\usepackage{subcaption}
\usepackage{sectsty}
\usepackage{titlesec}
\titleformat{\section}[hang]{\centering \bfseries\large}{\thesection.}{0.4em}{}

\renewcommand{\P}{\mathbb{P}}
\newcommand{\E}{\mathbb{E}}

\newtheorem{theorem}{Theorem}[section]

\DeclareMathOperator*{\argmax}{arg\,max}

\theoremstyle{definition}

\theoremstyle{definition}
\newtheorem{definition}[theorem]{Definition}

\def\one{{\bf 1}}

\def\bSigma{{\bf \Sigma}}

\def\w{{\bf w}}
\def\A{{\bf A}}

\def\X{{\bf X}}
\def\Y{{\bf Y}}

\def\Beta{\boldsymbol{\beta}}

\def\I{{\bf I}}
\def\F{{\cal F}}

\def\P{{\mathbb P}}
\def\E{{\mathbb E}}
\def\Var{{\rm Var}\,}
\def\Cov{{\rm Cov}\,}
\def\Corr{{\rm Corr}\,}

\def\|{\, | \,}

\def\diag{\text{diag}}

\def\diag{\text{diag}}

\allowdisplaybreaks

\title{Combining and Extremizing Real-Valued Forecasts}
\author{\vspace{-1em}Ville A. Satop\"a\"a and Lyle H. Ungar\thanks{Ville A. Satop\"a\"a is a Doctoral Candidate, Department of Statistics, The Wharton School of the University of Pennsylvania, Philadelphia, PA 19104-6340 (e-mail: satopaa@wharton.upenn.edu); Lyle H. Ungar is a Computer Scientist, Department of Computer and Information Science, University of Pennsylvania, Philadelphia, PA 19104-6309 (e-mail: ungar@cis.upenn.edu). This research was supported by a research contract to the University
of Pennsylvania and the University of California from the Intelligence
Advanced Research Projects Activity (IARPA) via the Department of
Interior National Business Center contract number D11PC20061. The
U.S. Government is authorized to reproduce and distribute reprints for
Government purposes notwithstanding any copyright annotation
thereon. Disclaimer: The views and conclusions expressed herein are
those of the authors and should not be interpreted as necessarily
representing the official policies or endorsements, either expressed
or implied, of IARPA, DoI/NBC, or the U.S. Government.}} 
\date{\vspace{-6.5ex}}

\begin{document}
\maketitle

\begin{abstract}
\singlespace
The weighted average is by far the most popular approach to combining
multiple forecasts of some future outcome. This paper shows that
both for probability or real-valued forecasts, a non-trivial weighted average
of different forecasts is always sub-optimal. More specifically, it is
not consistent with any set of information about the future outcome
even if the individual forecasts are. Furthermore, weighted averaging
does not behave as if it collects information from the forecasters and
hence needs to be \textit{extremized}, that is, 
systematically transformed away from the marginal mean. This paper
proposes a linear extremization technique for improving the weighted
average of real-valued forecasts. The resulting more extreme
version of the weighted average exhibits many properties of optimal
aggregation. Both this and the sub-optimality of the weighted average are illustrated with simple examples involving 
synthetic and real-world data. 
\end{abstract}

\newpage

\section{INTRODUCTION} \label{introduction}

Policy-makers often consult human or/and machine agents for forecasts
of some future outcome. For instance, multiple economics experts may
provide quarterly predictions of gross domestic product (GDP). Typically it is not possible to determine ex-ante which expert will be the most accurate, and even if this could be done, heeding only the most accurate expert's advice would ignore a potentially large amount of relevant information that is being contributed by the rest of the experts. Therefore a better alternative is to combine the
forecasts into a single consensus forecast that represents all the experts' advice. 
The policy-makers, however, can choose to aggregate the forecasts in many different ways. The final choice of the combination rule is crucial because it often decides how much of the experts' total information is incorporated and hence how well the consensus forecast performs in terms of predictive accuracy.

Possibly because of its simplicity and intuitive appeal, the most
popular approach to combining forecasts is the weighted average, sometimes also known as the linear opinion pool. 
This technique has a long tradition, with many empirical studies attesting to its benefits (see, e.g., \citealt{bates1969combination, clemen1989combining, armstrong2}). 
Even though the average forecast does not always outperform the best single forecaster \citep{hibon2005combine}, it is still considered state-of-the-art \citep{elliott2013handbook}
 in many fields, including economics \citep{blix2001good},  weather forecasting \citep{raftery2005using}, political science \citep{graefea2014combining}, and many others. In this paper, however, we show that non-trivial weighted averaging is suboptimal, and propose a simple transformation to improve it. A more detailed description of the contributions is given below.
%

In practice forecasts are typically either real-valued or probabilities of binary events, such as rain or no
rain tomorrow. \cite{Ranjan08} focus on the latter and
 explain how the quality of a probability forecast
(individual or aggregate) is typically measured in terms of
\textit{reliability} and \textit{resolution} (sometimes also known as calibration and
sharpness, respectively). Reliability describes how closely the
conditional event frequencies align with the forecast
probabilities. Resolution, on the other hand, measures how far the
forecasts are from the naive baseline forecast, that is, the marginal
event frequency. A forecast that is reliable and highly resolute is
very useful to the policy-maker because it is both accurate and close
to the most confident values of zero and one. Therefore a
well-established goal in probability forecasting is to maximize resolution subject to
reliability \citep{murphy1987general, gneiting2007probabilistic}.

Strikingly, \cite{Ranjan08} prove that any non-trivial weighted average
of two or more different, reliable probability forecasts is
unreliable and lacks resolution. In particular, they explain that such
a weighted average is  under-confident in a sense that it is overly
close to the marginal event frequency. This result is an important
contribution to the probability forecasting literature in part
because it points out a dramatic shortcoming of methodology that is
used widely in practice. However, the authors neither provide a
principled way of addressing the shortcoming nor interpret potential causes of the under-confidence.


The first step towards addressing these issues and improving the general practice of aggregation  is to understand what is meant by principled aggregation. This topic was discussed by \cite{satopaamodeling2, satopaamodeling} who propose the \textit{partial information 
framework} as a general platform for modeling and combining forecasts. Under this framework, the outcome and the forecasts share a probability space but without any restrictions on their dependence structure. Any forecast heterogeneity  is assumed to stem purely from information available to the forecasters and how they decide to use it. For instance, forecasters studying the same (or different) articles about the state of the economy may use distinct parts of the information and hence report different predictions of the next quarter's GDP. 
Even though, to date, this framework has been mainly used for constructing new aggregators, it also offers an ideal environment for analyzing other, already existing, aggregation techniques. No previous work, however, has used it to study weighted averaging of probability or real-valued forecasts.

%
%



The first contribution of this paper
leaves the type of forecasts unspecified and analyzes the weighted average of any univariate forecasts under the partial information framework. The results are general and encompass both probability and real-valued forecasts. First,  the aforementioned result
in \cite{Ranjan08} is generalized to any type of univariate
forecasts.
This result shows, for instance, that any non-trivial
weighted average of reliable predictions about the next quarter's GDP
is both unreliable and under-confident. 
Second, some general properties of optimal aggregation are enumerated. This  leads to
an original point of view on forecast aggregation, general, yet intuitive, descriptions of well-known properties
such as reliability and resolution, and an introduction of a new property, called \textit{variance expansion}, that is associated with aggregators whose variance is
never less than the maximum variance among the individual
forecasts. Such aggregators are called \textit{expanding} and can be considered to collect information from the individual forecasters. Showing that a non-trivial weighted average is never expanding leads to a mathematically precise yet
easy-to-understand explanation of why weighted averages tend to be
under-confident. This reasoning suggests that
under-confidence is not unique to the class of weighted averages but
extends to many other measures of central tendency, such as the
median, that also tend to reduce variance.


In probability forecasting the under-confidence of a simple aggregator, such as the average or median, is typically alleviated by a heuristic known as \textit{extremizing}, that is, by systematically transforming the aggregate towards its nearer extreme (at zero or one). 
For instance, \cite{Ranjan08} propose a beta transformation that extremizes the weighted average of
  the probability forecasts; \cite{satopaa} use a logistic regression model to extremize the average log-odds of the
  forecasts; many others, including  \cite{shlomi2010subjective}, \cite{baron2014two}, and  \cite{mellers2014psychological}, have also discussed extremization of probability forecasts.
Intuitively, extremization increases confidence by explicitly moving the aggregate closer to  the most confident values of zero and one. 
Naturally, the same intuition applies to probability forecasts of any categorical outcome. 
    However, if the outcome and forecasts are real-valued, it is not clear anymore what values represent the most confident forecasts. Consequently, it seems that extremization, as described above, lacks direction and cannot be applied. 
%
Furthermore, the idea of extremizing may seem counter-intuitive given the large amount of literature attesting to the benefits of shrinkage \citep{james1961estimation}.  These may be the main reasons why, to the best of our knowledge, no previous literature has discussed extremization of real-valued forecasts. 

   Therefore it is perhaps somewhat surprising that our second contribution shows that extremizing can improve aggregation also when the individual forecasts are real-valued. First, the notion of extremizing is made precise. This involves introducing a general definition that differs slightly from the above heuristic.    
        In particular, extremizing is redefined as a shift away from the least confident forecast, namely the marginal mean of the outcome, instead of towards the most confident (potentially undefined) values.  
%
%
Second, our definition and theoretical analysis motivate a convex optimization procedure that linearly extremizes the optimally weighted average of real-valued forecasts.
  The technique is illustrated on simple examples involving both synthetic and real-world data. In each example extremizing leads to improved aggregation with many of the optimal properties enumerated in the beginning of the analysis.

The rest of the paper is structured as follows. Section \ref{propertiesS} briefly
introduces the general partial information framework and discusses some properties of
the optimal aggregation within that framework. The class of weighted
averages is then analyzed in the light of these
properties. Section \ref{extremization} describes the optimization technique for extremizing the weighted average of
real-valued forecasts. Section \ref{simulation} illustrates this
technique and our theoretical results over synthetic
data. Section \ref{application} repeats the analysis over real-world data. The final section
concludes and discusses future research directions.

\section{FORECAST AND AGGREGATION PROPERTIES} \label{propertiesS}
\subsection{Optimal Aggregation}

Consider $N$ forecasters and suppose forecaster $j$ predicts $X_j$ for
some (integrable) quantity of interest $Y$.  The partial information
framework assumes that $Y$ and $X_j$, for $j = 1, \dots, N$, are
measurable random variables under some common probability space
$(\Omega, \F , \P)$.
Akin to \cite{murphy1987general}, \cite{Ranjan08}, \cite{jolliffe2012forecast}, and many others, 
 the forecasters are assumed to be \textit{reliable}, that is, conditionally unbiased
such that $\E(Y | X_j) = X_j$ for all $j = 1, \dots, N$.  To interpret
this assumption, observe that the principal $\sigma$-field $\F$ holds
all possible information that can be known about $Y$. Each
reliable forecast $X_j$ then generates a sub-$\sigma$-field
$\sigma(X_j) := \F_j \subseteq \F$ such that $X_j
= \E(Y|\F_j)$. Conversely, suppose that $X_j = \E(Y|\F_j)$ for some
$\F_j \subseteq \F$, then
\begin{align*}
\E(Y | X_j) &= \E[\E(Y|X_j,\F_j)|X_j] = \E[\E(Y|\F_j)|X_j] = \E(X_j|X_j) = X_j.
\end{align*}
Therefore a forecast is reliable if and only if it represents the optimal use of some information set, that is, it is consistent with some partial information $\F_j \subseteq \F$. Given that at this level of specificity the framework is highly general and hence likely to be a good approximation of real-world prediction polling, it offers an ideal platform for analyzing different aggregators. 

In this paper an aggregator is defined to be any forecast that is measurable with respect to $\F'' := \sigma(X_1, \dots, X_N)$, namely the $\sigma$-field generated by the individual forecasts. For the sake of notational clarity, aggregators are denoted with different versions of the script symbol $\mathcal{X}$. If $\E\left(Y^2\right) < \infty$, the conditional expectation
$\mathcal{X}'' := \E(Y | \F'')$ minimizes the expected quadratic loss
among all aggregators (see, e.g., \citealt{durrett2010probability}). This forecast is called the \textit{revealed
aggregator} because it optimally utilizes all the information that the
forecasters' reveal through their forecasts. Even though
$\mathcal{X}'' $ is typically too abstract to be applied in practice,
it provides an optimal baseline for aggregation
efficiency. Therefore studying its properties gives guidance for
improving aggregators currently used in practice. Some of these
properties are summarized in the following theorem. The proof is deferred to the Appendix.

\begin{theorem} \label{optimal}
Suppose that $X_j = \E(Y | X_j)$ for all $j = 1, \dots, N$ and denote the revealed aggregator with $\mathcal{X}'' = \E(Y | \F'')$, where $\F'' = \sigma(X_1, \dots, X_N)$. 
Let $\delta_{max} := \max_j \{ \Var(X_j)  \}$ be the maximal variance among the individual forecast.
 Then the following holds.
\begin{enumerate}[i)] \label{properties}
\item \textbf{Marginal Consistency.} $\mathcal{X}''$ is marginally consistent:  $\E(\mathcal{X}'') = \E(Y) :=  \mu_0$.
\item \textbf{Reliability.} $\mathcal{X}''$ is reliable: $\E(Y|\mathcal{X}'') = \mathcal{X}''$. 
\item \textbf{Variance Expansion.} $\mathcal{X}''$ is expanding: $\delta_{max} \leq \Var(\mathcal{X}'')$. In words, the variance of $\mathcal{X}''$ is always at least as large as that of the most variable forecast. 
\end{enumerate}
\end{theorem}
Marginal consistency states that the forecast and the outcome agree in expectation. If $X_j$ is reliable, then $\E(X_j) = \E[\E(Y|X_j)] = \E(Y) = \mu_0$. Consequently, all reliable forecasts (individual or aggregate) are marginally consistent. The converse, however, is not true. For instance, Theorem \ref{contraction} (see Section \ref{contraction})  shows that any non-trivial weighted average is marginally consistent but unreliable. This is an important observation because it provides a technique for proving lack of reliability via marginal inconsistency -- a task that is generally much easier than disproving reliability directly.

Given that each reliable forecast can be associated with a sub-$\sigma$-field and  that conditional expectation is a contraction in $L^2$ \citep[Theorem 5.1.4.]{durrett2010probability}, the variance of any reliable forecast (individual or aggregate) is always upper-bounded by $\Var(Y)$. 
Theorem \ref{optimal} further shows that the corresponding lower bound for $\Var(\mathcal{X}'')$ is the maximum variance among the forecasters. To interpret this lower bound, consider 
an increasing sequence of $\sigma$-fields $\F_0 = \{\emptyset, \Omega\} \subseteq \F_1 \subseteq \dots \subseteq \F_R \subseteq \F$ and the corresponding forecasts $X_r = \E(Y|\F_r)$ for $r = 0,1, \dots, R$. According to \citet[Proposition 2.1]{satopaamodeling2}, the variances of these forecasts respect the same order as their information sets: $\Var(X_0) \leq \Var(X_1) \leq \dots \leq \Var(X_R) \leq \Var(Y)$. This suggests that the amount of information used in a reliable forecast is reflected in its variance. Naturally, if an aggregator collects information from a group of forecasters, it should use at least as much information as the most informed individual forecaster; that is, its variance should exceed that of the individual forecasters'. Therefore any aggregator that \textit{expands} variance and satisfies this condition is
considered a collector of information.

Recall that in probability forecasting a well-established goal is to maximize resolution subject to reliability. This goal can be easily interpreted intuitively with the help of partial information. First, conditioning on reliability requires the forecast to be consistent with some set of information about $Y$. Maximizing the resolution of this forecast takes it as far from $\mu_0$ as possible. This is equivalent to increasing the variance of the forecast as close to the theoretical upper bound $\Var(Y)$ as possible. Therefore the goal is equivalent to maximizing the amount of information that the forecast is consistent with. Intuitively, this is very reasonable and should be considered as the general goal in forecasting.

\subsection{Weighted Averaging} \label{contraction}

The rest of the paper analyzes the most commonly used aggregator, namely the weighted average. The following theorem shows that a non-trivial weighted average is neither expanding nor reliable and therefore can be considered suboptimal. The proof is again deferred to the Appendix. A similar result does not hold for all linear combinations of the individual forecasts. For instance, Section \ref{simulation} describes a model under which the optimal aggregator $\mathcal{X}''$ is always a linear combination of the individual $X_j$'s.

\begin{theorem}\label{contraction}
Suppose that $X_j = \E(Y | X_j)$ for $j = 1, \dots, N$. Denote the weighted average with $\mathcal{X}_w := \sum_{j=1}^N w_jX_j$, where  $w_j \geq 0$, for all $j = 1, \dots, N$, and $\sum_{j=1}^N w_j = 1$.  Let $m = \argmax_j \{ \Var(X_j)  \}$ identify the forecast with the maximal variance $\delta_{max} = Var(X_m)$. Then the following holds.
\begin{enumerate}[i)]
\item  $\mathcal{X}_w$ is marginally consistent.

\item $\mathcal{X}_w$ is not reliable, that is, $\P\left[\E(Y | \mathcal{X}_w) \neq \mathcal{X}_w\right] > 0$ if there exists a forecast pair $i \neq j$ such that $\P(X_i \neq X_j) > 0$ and $w_i, w_j > 0$. In words, $\mathcal{X}_w$ is necessarily unreliable if it assigns positive weight to at least two different forecasts. 
 
 \item Under the conditions of item ii), $\mathcal{X}_w$ lacks resolution. More specifically, if $\mathcal{X}_w' :=  \E(Y| \mathcal{X}_w)$ is the reliable version of $\mathcal{X}_w$, then $\E(\mathcal{X}_w) = \E(\mathcal{X}_w') = \mu_0$ but $\Var(\mathcal{X}_w) < \Var(\mathcal{X}_w')$. In other words, $\mathcal{X}_w$ is under-confident in a sense that it is closer to the marginal mean $\mu_0$ than its reliable version $\mathcal{X}_w'$. \label{underconfA}
 
\item $\mathcal{X}_w$ is not expanding. In particular, $\Var(\mathcal{X}_w) \leq \delta_{max}$, which shows that $\mathcal{X}_w$ is under-confident in a sense that it is as close or closer to the marginal mean $\mu_0$ than the revealed aggregator $\mathcal{X}''$.  Furthermore, $\Var(\mathcal{X}_w) = \Var(\mathcal{X}'')$ if and only if both $\mathcal{X}_w = \mathcal{X}'' = X_m$; that is,  $X_m$ provides all the information necessary for $\mathcal{X}''$, and $\mathcal{X}_w$ assigns all weight to $X_m$ (or to a group of forecasts all equal to $X_m$). \label{underconfB}
\end{enumerate}
\end{theorem}

This theorem discusses under-confidence under two different baselines. Item \ref{underconfA}) is a generalization of \citet[Theorem 2.1.]{Ranjan08}. Intuitively, it states that if $\mathcal{X}_w$ is trained to use its information accurately, the resulting aggregator is more confident. Therefore under-confidence is defined relative to the reliable version of  $\mathcal{X}_w$. Under this kind of comparison, however, a reliable aggregator is never under-confident. For instance, an aggregator that ignores the individual forecasts and always returns the marginal mean $\mu_0$ is reliable and hence would not be considered under-confident. Intuitively, however, it is clear that no aggregate forecast is more under-confident than the marginal mean $\mu_0$. To address this drawback, item \ref{underconfB}) defines under-confidence relative to  the revealed aggregator instead. Such a comparison estimates whether the weighted average is as confident as it should be given the information it received through the forecasts. Item \ref{underconfB}) shows that this happens only if all the weight is assigned to a forecaster whose information set contains every other forecasters' information. However, even if $\mathcal{X}_w$ could pick out the most informed forecaster ex-ante, the chances of a single forecaster knowing everything that the rest of the forecasters know is extremely small in practice. In essentially all other cases, $\mathcal{X}_w$ is under-confident, unreliable,  
and hence not consistent with some set of information about $Y$. 

Unfortunately, this shortcoming spans across all measures of central tendency. These aggregators reduce variance and hence are separated from the revealed aggregator by the maximum variance among the individual forecasts. For instance, \cite{papadatos1995maximum} discuss the maximum variance of different order statistics and show that the variance of the median is upper bounded by the global variance of the individual forecasts. Given that such aggregators are not expanding, they cannot be considered to collect information. To illustrate, consider a group of forecasters, each independently making a probability forecast of $0.9$ for the occurrence of some future event. If these forecasters are using different evidence, then clearly the combined evidence should give an aggregate forecast somewhat greater than $0.9$. In this simple scenario, however, measures of central tendency will always aggregate to $0.9$. Therefore they fail to account for the information heterogeneity among the forecasters. 
Instead, they reduce ``measurement error,'' which is philosophically very different to the idea of information aggregation discussed in this paper. 


%


Theorem \ref{contraction}, however, is not only negative in nature; it is also  constructive in several different ways. First, 
it motivates a general and precise definition of extremizing:
\begin{definition} \label{extrem}
\textbf{Extremization.}
Consider two reliable forecasts $X_i$ and $X_j$. Denote their common marginal mean with $\E(X_i) = \E(X_j) = \mu_0$. The forecast $X_j$ \textit{extremizes} $X_i$ if and only if either $X_j \leq X_i \leq \mu_0$ or $\mu_0 \leq X_i \leq X_j $ always holds.
\end{definition}

\noindent
It is interesting to contrast this definition with the popular extremization heuristic in the context of probability forecasting.  Definition \ref{extrem} suggests that simply moving, say, the average probability forecast closer to zero or one improves the aggregate if and only if the marginal probability of success is $0.5$. In other cases naively following the heuristic may end up degrading the aggregate. For instance, consider a geographical region where rain is known to occur on $20$\% of the days. If the average probability forecast of rain tomorrow is $0.30$, instead of following the heuristic and shifting this aggregate towards zero and hence closer to the marginal mean of $0.20$, the aggregate should be actually shifted in the opposite direction, namely closer to one. Second, Theorem \ref{contraction} suggests that extremization, as defined formally above, is likely to improve the weighted average of any type of univariate forecasts. This justifies the construction of a broader class of extremizing techniques. In particular, the second part of item \ref{underconfB}) states that extremizing is likely to improve the weighted average when 
 the single most informed forecaster knows a lot less than all the forecasters know as a group. To illustrate this, the next section introduces a simple optimization procedure that extremizes the weighted average of real-valued forecasts.
%
%

\section{EXTREMIZING REAL-VALUED FORECASTS} \label{extremization}


Estimating the weights and the amount of extremization requires the forecasters to address more than one related problems. For instance, they may participate in separate yet similar prediction problems or give repeated forecasts on a single recurring event. Across such problems the weights and the resulting under-confidence are likely to remain stable, allowing the aggregator parameters to be estimated based on multiple predictions per forecaster. Therefore, from now on, suppose that the forecasters address $K \geq 2$ problems. Denote the outcome of the $k$th problem with $Y_k \in \mathbb{R}$ and let $X_{jk} \in \mathbb{R}$ represent the $j$th forecaster's prediction for this outcome. 

Extremization requires at least two parameters: the marginal mean, which acts as the pivot point and decides the direction of extremizing, and the amount of extremization itself. Extremization, of course, could be performed in many different ways. However, if $\mathcal{X}_k^*$ denotes the extremized version of the weighted average for the $k$th problem, then probably the simplest and most natural starting point is the following:
\begin{align*}
\mathcal{X}_k^* = \alpha  \left(  \w'\X_{k} - \mu_0\right) + \mu_0,
\end{align*}
 where $\X_k = (X_{1k}, \dots, X_{Nk})'$ collects the forecasts for the $k$th outcome, $\w = (w_1, \dots, w_N)'$ is the weight vector, and $\alpha \in (1, \infty)$ (or $\alpha \in [0, 1)$) leads to extremization (or contraction towards $\mu_0$, respectively). If $\alpha = 1$, then $\mathcal{X}^*$ is equal to the weighted average $\mathcal{X}_w$.  This linear form is particularly convenient because it leads to efficient parameter estimation and also maintains marginal consistency of $\mathcal{X}_w$; that is, $\E(\mathcal{X}^*) = \mu_0$ for all values of $\alpha$.  However, $\Var(\mathcal{X}^*)$ increases in $\alpha$ such that $\Var(\mathcal{X}^*) = \alpha^2 \Var(\mathcal{X}_w) > \Var(\mathcal{X}_w)$ for all $\alpha > 1$. Therefore, for a large enough $\alpha$, $\mathcal{X}^*$ is both marginally consistent and expanding. These properties hold even if the weighted average is replaced by some other marginally consistent aggregator. However, given that the main purpose of this procedure is to illustrate Theorem \ref{contraction}, this paper only considers the weighted average.

Recall that the forecasts are assumed calibrated and hence marginally consistent with the outcomes. Therefore an unbiased estimator of the prior mean $\mu_0$ is given by the average of the forecasts $\frac{1}{NK} \sum_{k=1}^K\sum_{j=1}^N X_{jk}$ or, alternatively, by the average of the outcomes $\frac{1}{K} \sum_{k=1}^K Y_k$. Estimating $\mu_0$ in this manner, however, leads to a two-step estimation procedure. A more direct approach is to estimate all the parameters, namely $\alpha$, $\mu_0$, and $\w$, jointly over some criterion. If $Y_k$ has an explicit likelihood in terms of $\mathcal{X}^*$, then the parameters can be estimated by maximizing this likelihood. Assuming an explicit parametric form, however, can be avoided by recalling from Section \ref{contraction} that the revealed aggregator $\mathcal{X}''$ utilizes the forecasters' information optimally and minimizes the expected quadratic loss among all functions measurable with respect to $\F''$. Ideally, $\mathcal{X}^*$ would behave similarly to $\mathcal{X}''$. Therefore it makes sense to estimate its parameters by minimizing the average quadratic loss over some training set. Section \ref{simulation} shows that this is likely to improve both the resolution and reliability of the weighted average. 

These considerations lead to the following estimation problem:
\begin{align}
 \label{firstProblem}
 \begin{split}
\text{minimize } & \sum_{k=1}^K \left[ \alpha  \left(  \w'\X_{k} - \mu_0\right) + \mu_0 - Y_k \right]^2\\
\text{subject to } & w_j \geq 0 \text{ for } j = 1, \dots, N,\\
& \sum_{j=1}^N w_j = 1, \text{ and }\\
& \alpha \geq 0.
\end{split}
\end{align}
To express this problem in a form that is more amenable to estimation, denote an $N \times N$ identity matrix with $\I_N$,  a vector of $K$ ones with $\one_K$, and a vector of $N$ zeros with $\boldsymbol{0}_N$. If $\Y = (Y_1, \dots, Y_K)'$, $\X = (\one_K, (\X_1, \dots, \X_K)')$, and $\A = (\boldsymbol{0}_N, \I_N)$, then  problem (\ref{firstProblem}) is equivalent to
\begin{align}
 \label{secondProblem}
 \begin{split}
\text{minimize } & \frac{1}{2} \Beta' \X'\X \Beta - \Y'\X\Beta\\
\text{subject to } & -\A\Beta \leq \boldsymbol{0}_N,
\end{split}
\end{align}
where the inequality is interpreted element-wise and $\Beta$ is a vector of $N+1$ optimization parameters. Given that $\X'\X$ is always positive semidefinite, problem (\ref{secondProblem}) is a convex quadratic program that can be solved efficiently with standard optimization techniques. 
 If $\Beta^* = (\beta_0^*, \dots, \beta_{N}^*)'$ represents the solution to (\ref{secondProblem}), the optimal values of the original parameters can be recovered by
\begin{align*}
\alpha^* &= \sum_{j=1}^N \beta_j^*,\\
w_j^* &=  \beta_j^*/\alpha^* \text{ for } j = 1, \dots, N, \text{ and}\\
\mu_0^* &= -\beta_0^*/(1-\alpha^*).
\end{align*}
The next two sections apply and evaluate this method both on simulated and real-world data. 

\section{SIMULATION STUDY} \label{simulation}

This section illustrates Theorem \ref{contraction} on data generated from the Gaussian partial information model introduced in \cite{satopaamodeling2, satopaamodeling}  as a close yet practical specification of the general partial information framework.
%
 The simplest version of this model occurs when the outcome $Y$ and the forecasts $X_j$ are real-valued with mean zero. The observables for the $k$th problem are then generated jointly from the following multivariate Gaussian distribution:
\begin{align}
\left(\begin{matrix} Y_k \\ X_{1k}\\ \vdots \\ X_{Nk} \end{matrix}\right) &\sim \mathcal{N}_{N+1}\left( 
 \boldsymbol{0}, \left(\begin{matrix} 
1 & \diag(\bSigma)'\\
\diag(\bSigma) &\bSigma\\
 \end{matrix}\right) 
 :=
 \left(\begin{array}{c | c c cc }
1 & \delta_1 & \delta_2 & \dots & \delta_N  \\ \hline
\delta_1 & \delta_1 &\rho_{1,2} & \dots & \rho_{1,N}   \\ 
\delta_2 & \rho_{2,1} & \delta_2 & \dots & \rho_{2,N}  \\ 
\vdots & \vdots & \vdots & \ddots & \vdots  \\ 
\delta_N & \rho_{N,1} & \rho_{N,2} & \dots & \delta_N\\ 
 \end{array}\right)\right),  \label{NExperts}
\end{align}
where the covariance matrix describes the \textit{information structure} among the forecasters. In particular, the maximum amount of information is $1.0$. The diagonal entry $\delta_j \in [0,1]$ represents the amount of information used by forecaster $j$ such that if $\delta_j = 1$ (or $\delta_j = 0$), the forecaster always reports the correct answer $Y_k$ (or the marginal mean $\mu_0 = 0$, respectively). The off-diagonal $\rho_{i,j}$, on the other hand, can be regarded as the amount of information overlap between forecasters $i$ and $j$. Using the  well-known properties of a conditional multivariate Gaussian distribution, \cite{satopaamodeling2, satopaamodeling} show that under this model the forecasts are reliable
%
and that the revealed aggregator for the $k$th problem is  $\mathcal{X}_k''  = \E(Y_k | \X_k) = \diag(\bSigma)'\bSigma^{-1}\X_k$. 



\begin{figure}[t!]
        \centering
        \begin{subfigure}[b]{0.49\textwidth}
                \includegraphics[width=\textwidth]{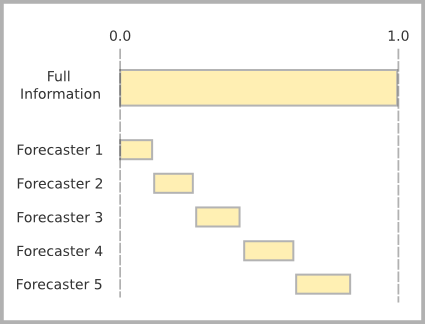}
                \caption{No Information Overlap}
                \label{DiagramsA}
        \end{subfigure}%
        ~ 
        \begin{subfigure}[b]{0.49\textwidth}
                \includegraphics[width=\textwidth]{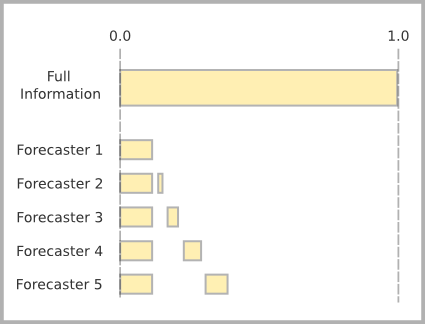}
                \caption{High Information Overlap}
                \label{DiagramsB}
        \end{subfigure}
   \caption{Information Distribution Among $N = 5$ Forecasters. The top bar next to \textit{Full Information} represents all possible information that can be known about $Y$. The bar leveled horizontally with \textit{Forecaster $j$} represents the information used by that forecaster.}    
        \label{Diagrams}
\end{figure}

The distribution (\ref{NExperts}) is particularly useful because it provides a realistic model for testing aggregation under different information structures. This section considers $N = 5$ forecasters under two different structures:
\begin{enumerate}
\item[] \textit{No Information Overlap.}  Fix $\delta_j = 0.1 + 0.02j$ for $j = 1, \dots, 5$ and let $\rho_{i,j} = 0$ for all $i,j$. Therefore the forecasters have independent information sources. This information structure is illustrated in Figure \ref{DiagramsA}. Summing up the individual variances shows that as a group the forecasters know $80\%$ of the total information. The revealed aggregator reduces to $\mathcal{X}_k'' = \sum_{j=1}^5 X_{jk}$, has variance $0.80$, and therefore efficiently uses all the forecasters' information.

\item[] \textit{High Information Overlap.} Fix $\delta_j = 0.1 + 0.02j$ for $j = 1, \dots, 5$ and let $\rho_{i,j} = 0.12$ for all $i,j$. Therefore the forecasters have significant information overlap and as a group know only $32\%$ of the total information. This information structure is illustrated in Figure \ref{DiagramsB}. The revealed aggregator reduces to $ \mathcal{X}_k'' = \left( \sum_{j=2}^5 X_{jk} \right) - 3X_{1k}$, has variance $0.32$, and therefore efficiently uses all the forecasters' information.
\end{enumerate}

The competing aggregators are the equally weighted average $\bar{\mathcal{X}}$, the optimally weighted average $\mathcal{X}_w$, the extremized version of the optimally weighted average $\mathcal{X}^*$, and the revealed aggregator $\mathcal{X}''$. The parameters in $\mathcal{X}^*$ and  $\mathcal{X}_w$ are first estimated by minimizing the average quadratic loss over a training set of $10,000$ draws from (\ref{NExperts}).  After this, all the competing aggregators are evaluated on an independent test set of another $10,000$ draws from (\ref{NExperts}). 
Therefore all the following results, apart from the parameter estimates, represent out-of-sample performance.

\begin{figure}[t!]
        \centering
        \begin{subfigure}[b]{0.240\textwidth}
                \includegraphics[width=\textwidth]{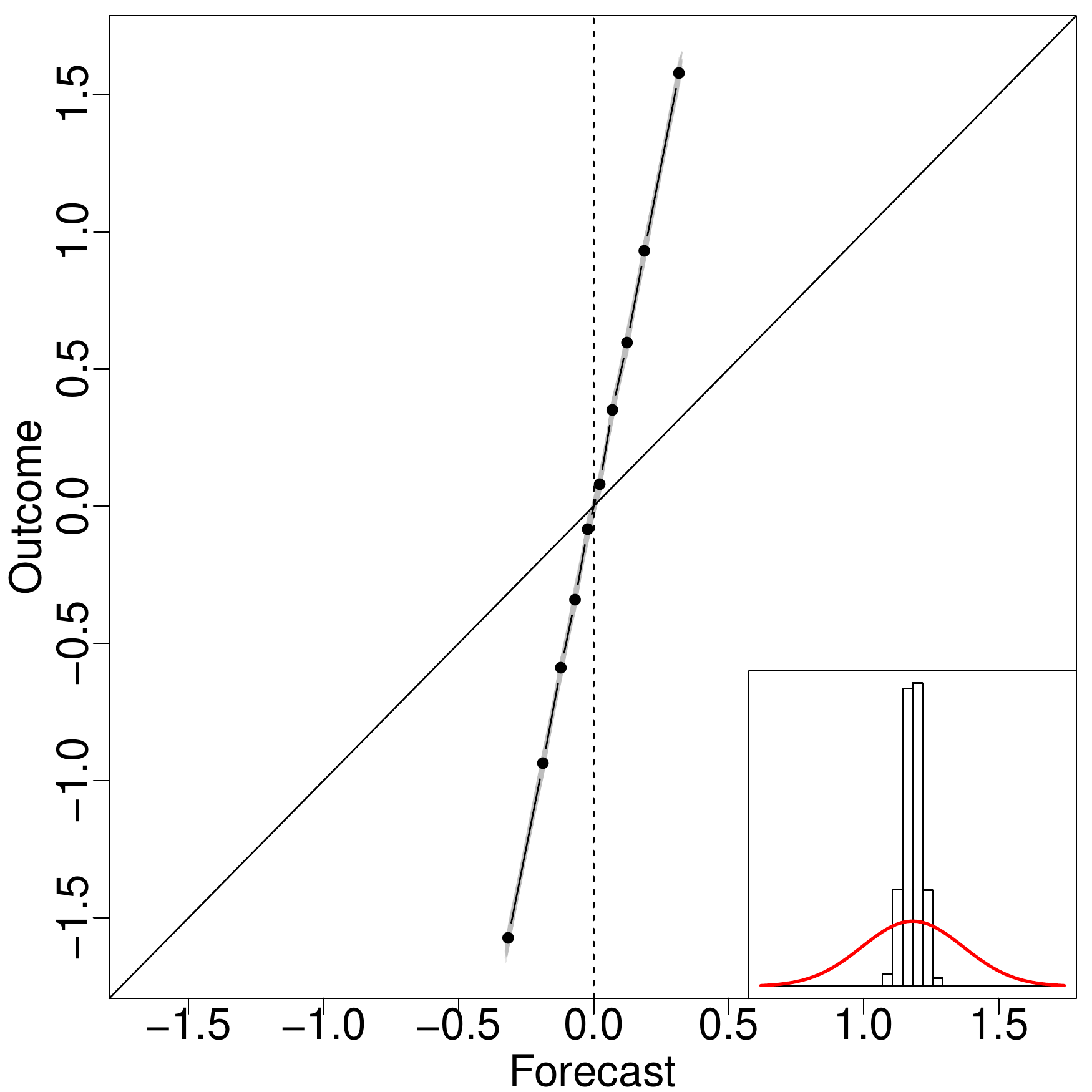}
                \caption{$\bar{\mathcal{X}}$}
        \label{RelEWANo}
        \end{subfigure}%
        ~ 
        \begin{subfigure}[b]{0.240\textwidth}
                \includegraphics[width=\textwidth]{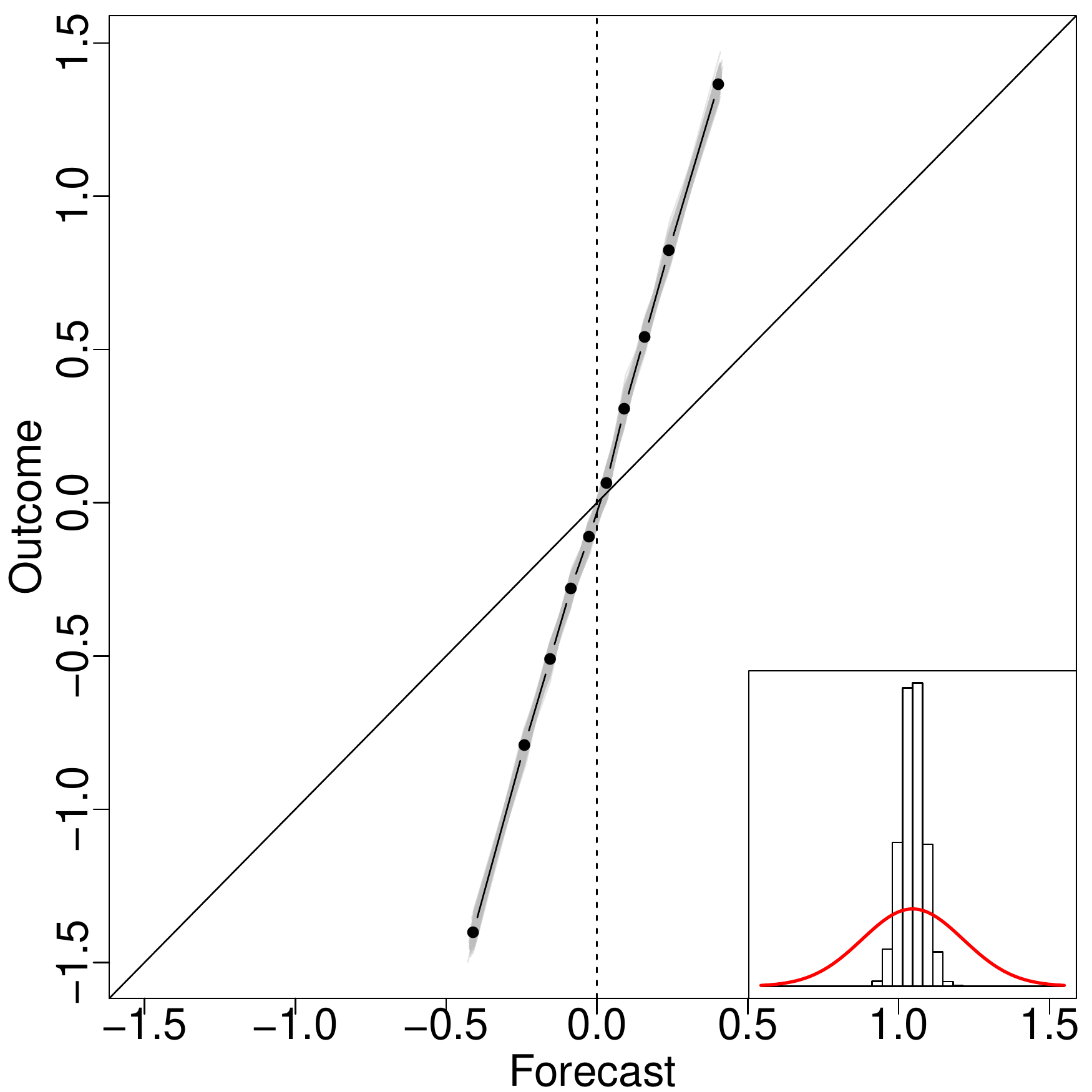}
                \caption{$\mathcal{X}_w$}
        \label{RelOWANo}
        \end{subfigure}
        ~ 
        \begin{subfigure}[b]{0.240\textwidth}
                \includegraphics[width=\textwidth]{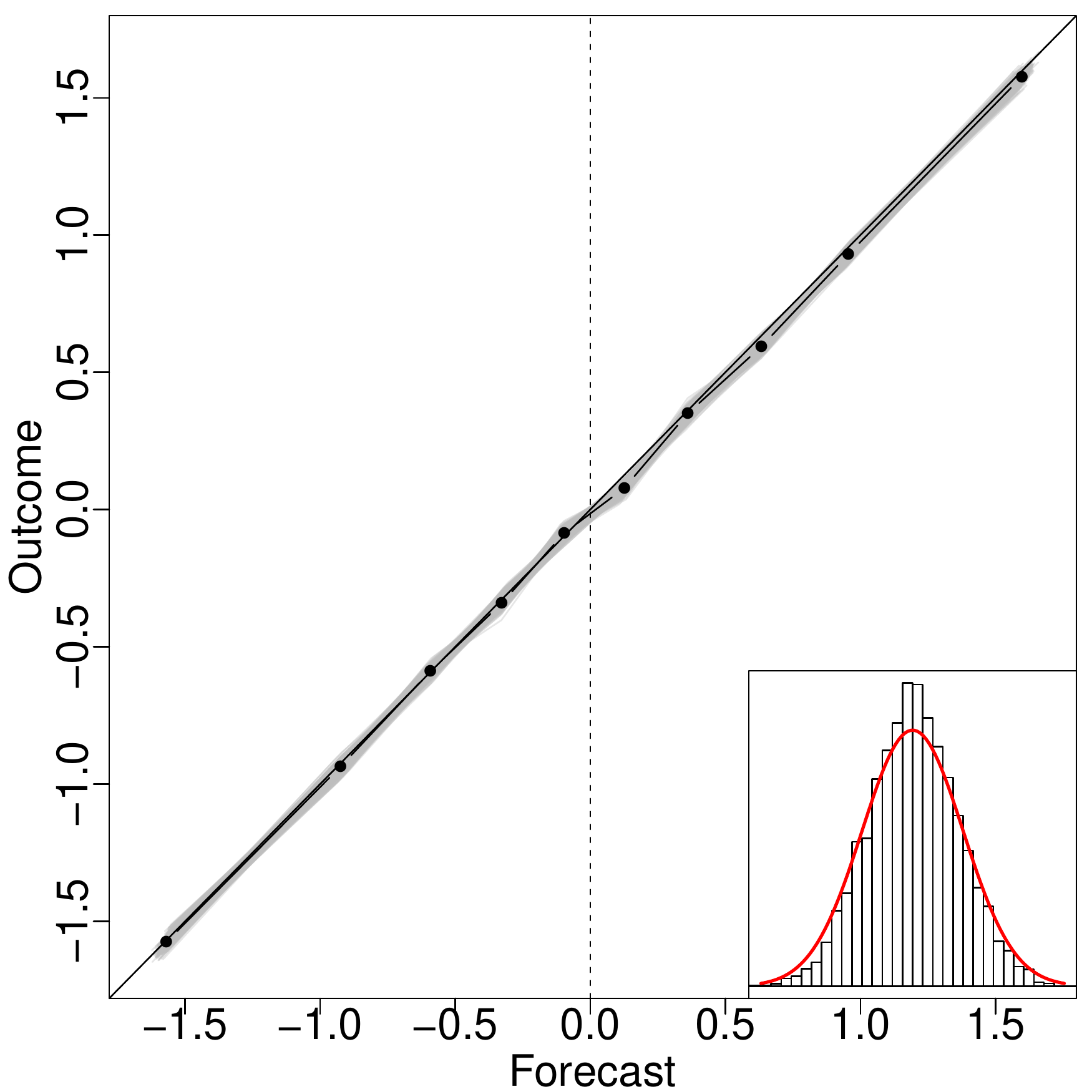}
                \caption{$\mathcal{X}^*$ }
        \label{ELOPNoOverlap}
        \end{subfigure}
                \begin{subfigure}[b]{0.240\textwidth}
                \includegraphics[width=\textwidth]{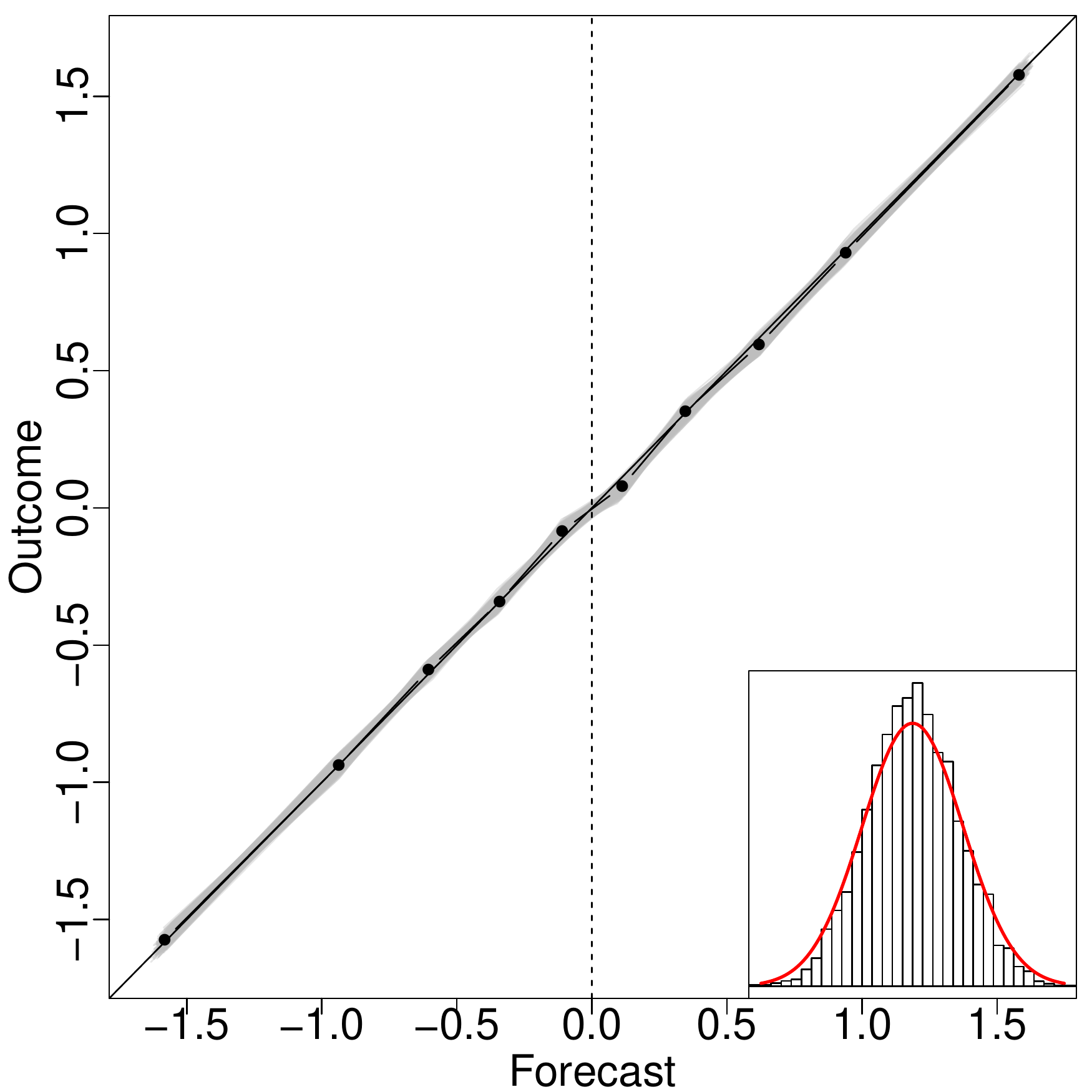}
                \caption{$\mathcal{X}''$ }
        \label{RevNoOverlap}
        \end{subfigure}
          \caption{Synthetic Data. Out-of-sample reliability under no information overlap. }
        \label{NOVerlap}
\end{figure}

\begin{figure}[t!]
        \centering
        \begin{subfigure}[b]{0.240\textwidth}
                \includegraphics[width=\textwidth]{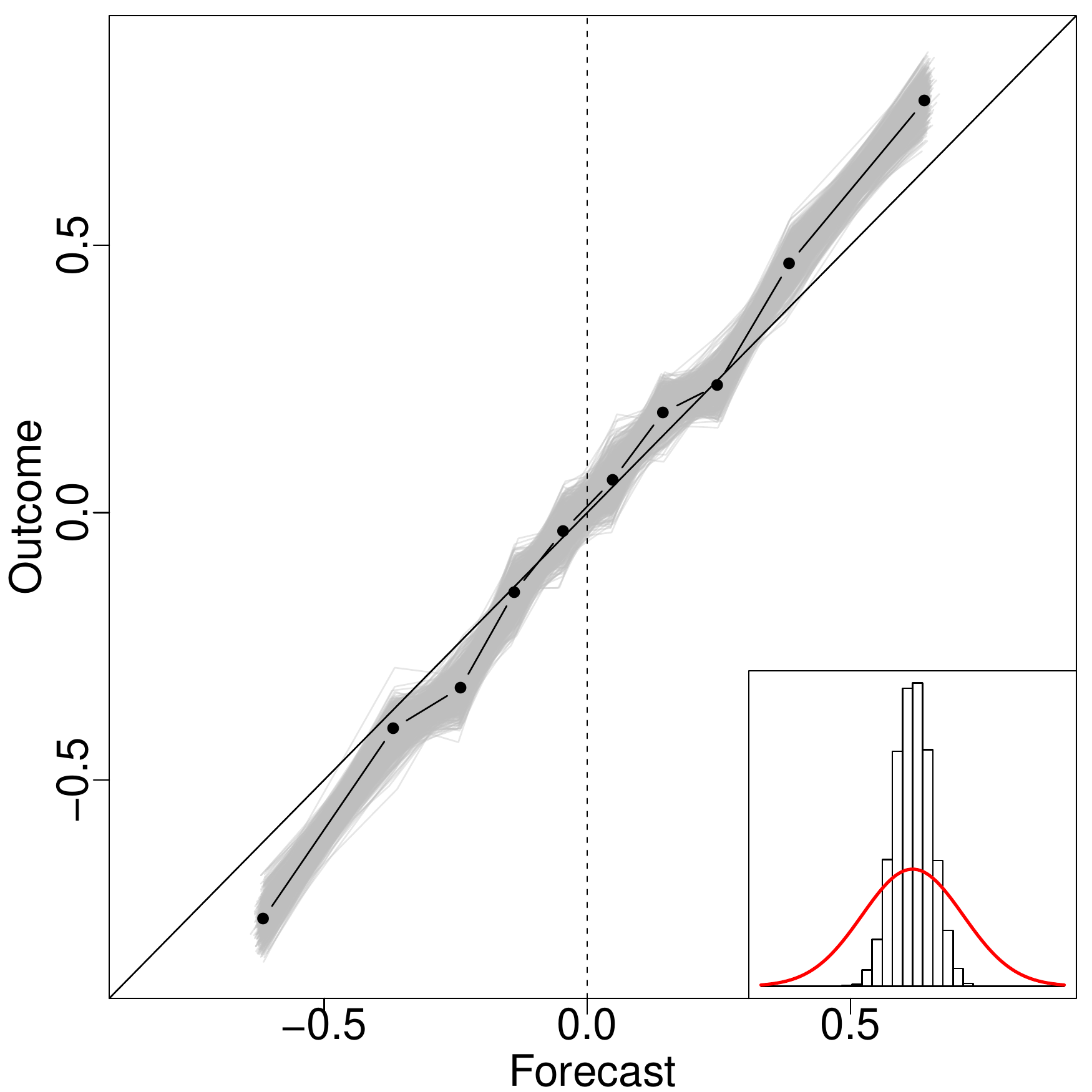}
                \caption{$\bar{\mathcal{X}}$}
        \label{RelEWAHigh}
        \end{subfigure}%
        ~ 
        \begin{subfigure}[b]{0.240\textwidth}
                \includegraphics[width=\textwidth]{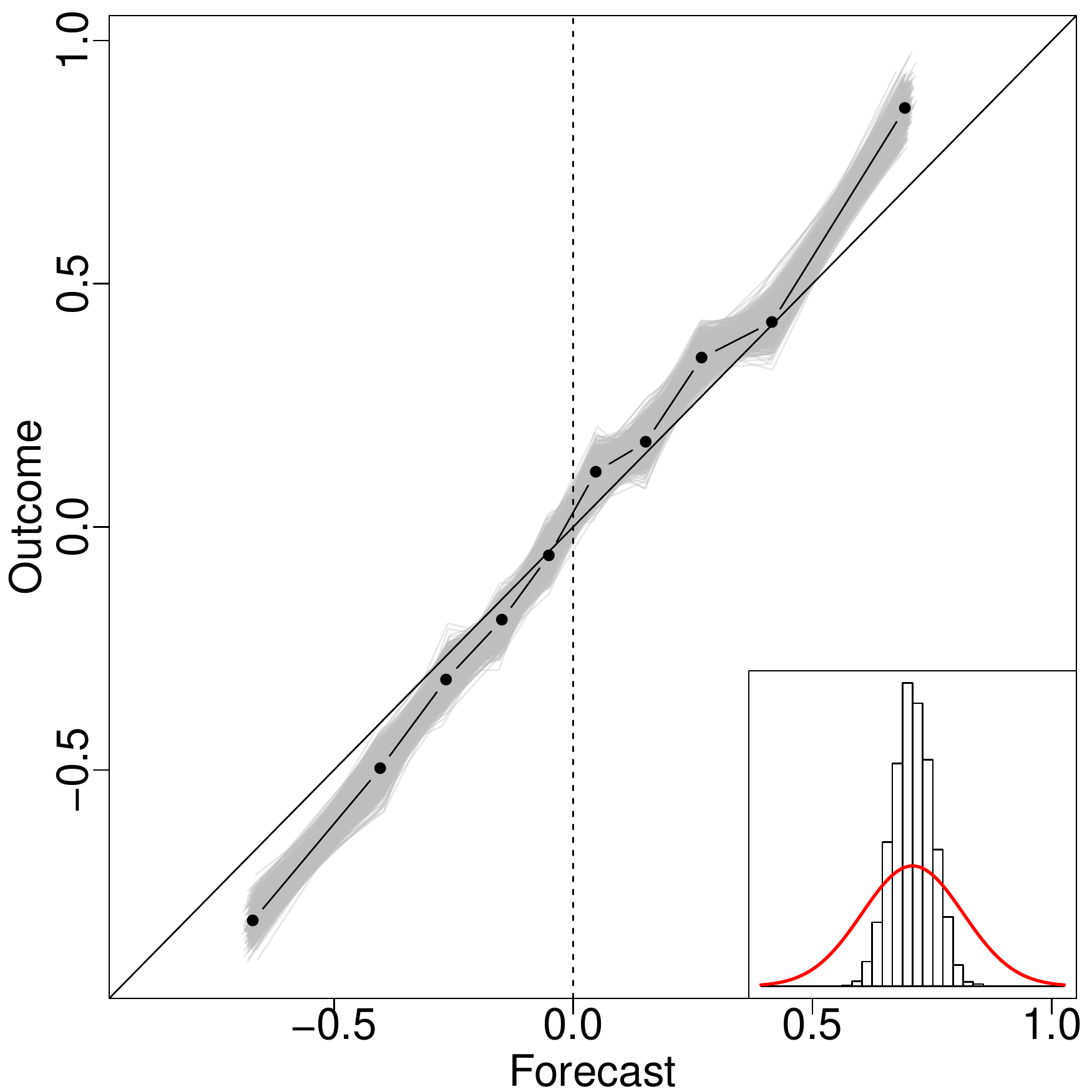}
                \caption{$\mathcal{X}_w$}
        \label{RelOWAHigh}
        \end{subfigure}
        ~ 
        \begin{subfigure}[b]{0.240\textwidth}
                \includegraphics[width=\textwidth]{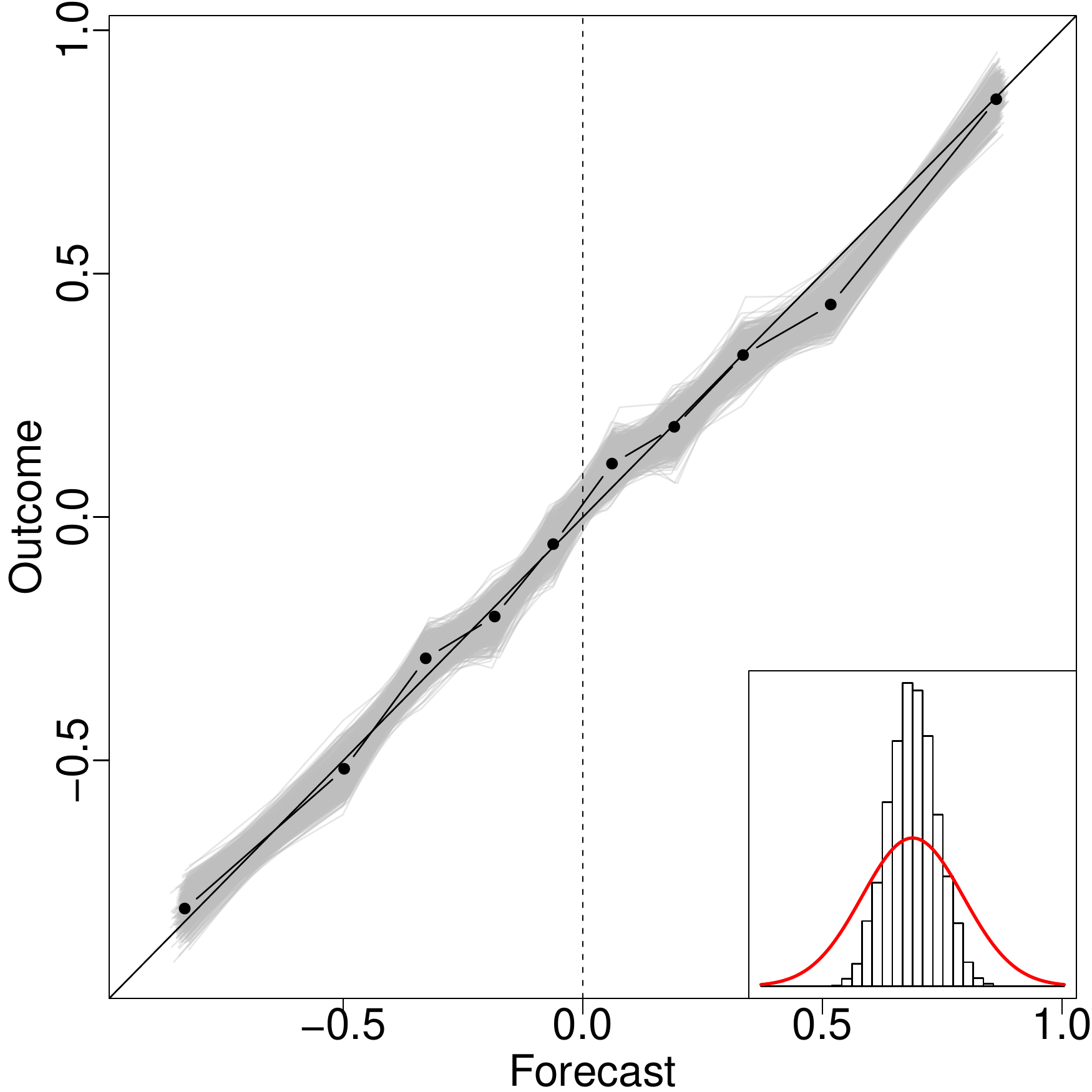}
                \caption{$\mathcal{X}^*$ }
        \label{ELOPHighOverlap}
        \end{subfigure}
           \begin{subfigure}[b]{0.240\textwidth}
                \includegraphics[width=\textwidth]{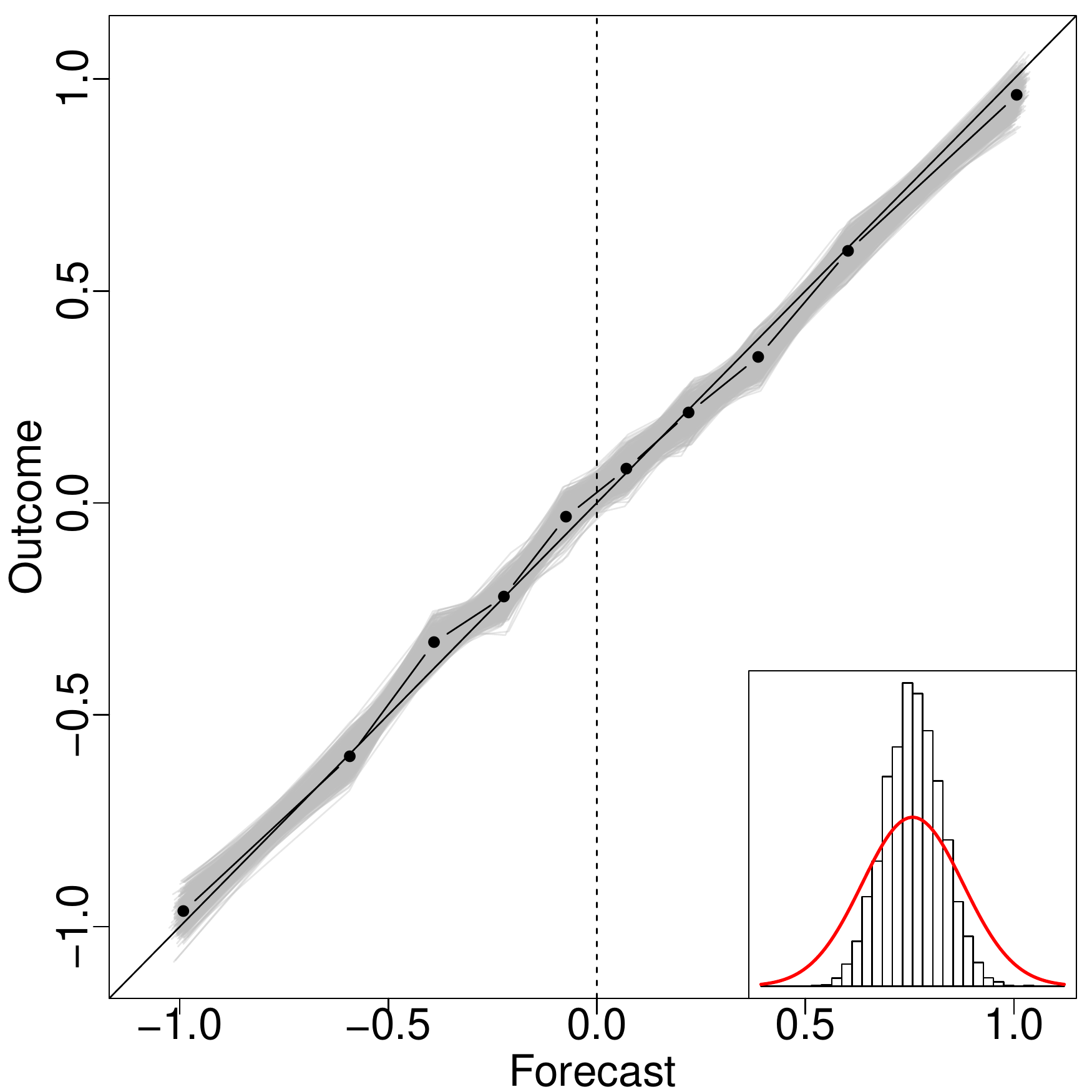}
                \caption{$\mathcal{X}''$ }
        \label{RevHighOverlap}
        \end{subfigure}
          \caption{Synthetic Data. Out-of-sample reliability under high information overlap. }
   \label{HighOverlap}
\end{figure}

In probability forecasting the quality of the predictions is typically assessed using a reliability diagram. The idea is to first sort the outcome-forecast pairs into some number of bins based on the forecasts and then plot the average forecast against the average outcome within each bin. Figures \ref{NOVerlap} and  \ref{HighOverlap}  generalize this to continuous outcomes by replacing the conditional empirical event frequency with the conditional average outcome. The bins are chosen so that they all contain the same number of forecast-outcome pairs. The vertical dashed line represents the marginal mean $\mu_0 = 0$.  The plots have been scaled such that the identity function shows as the diagonal. Any deviation from this diagonal suggests lack of reliability. The grey area represents the reliability diagrams of a $1,000$ bootstrap samples of the forecast-outcome pairs. Therefore it serves as a visual guide for assessing uncertainty. The inset histograms help to assess resolution by comparing the empirical distribution of the forecasts against the prior distribution of $Y$, namely the standard Gaussian distribution represented by the red curve. In particular, if the forecast is reliable, then the closer its empirical distribution is to the standard Gaussian, the more information is being used in the forecast. 

\begin{table}[t]
\centering
\caption{Synthetic Data. Estimated parameter values.}
\begin{tabular}{llrrrrrrr}
  \hline \hline
Scenario & Forecast & $\mu_0$ & $\alpha$ & $w_1$ & $w_2$& $w_3$& $w_4$& $w_5$ \\ 
  \hline
\multirow{2}{*}{No Overlap} & $\mathcal{X}_w$ &  &  & 0.0000 & 0.1080 & 0.2293 & 0.3025 & 0.3601 \\ 
&   $\mathcal{X}^*$ & 0.0004 & 5.0137 & 0.1964 & 0.2023 & 0.2008 & 0.2006 & 0.2000  \\  \rule{0pt}{2.9ex} 
\hspace{-0.2em}\multirow{2}{*}{High Overlap}  & $\mathcal{X}_w$ &  &  & 0.0000 & 0.0000 & 0.0440 & 0.4262 & 0.5298 \\ 
  & $\mathcal{X}^*$ & -0.0077 & 1.3048 & 0.0000 & 0.0000 & 0.1456 & 0.3959 & 0.4585 \\ 
   \hline
\end{tabular}
\label{NoParams}
\end{table}

Figures \ref{RevNoOverlap} and \ref{RevHighOverlap} present the reliability diagrams for $\mathcal{X}''$ under no and high information overlap, respectively. Comparing these plots to the corresponding reliability diagrams of $\bar{\mathcal{X}}$ and $\mathcal{X}_w$ in the same figures, reveals that $\bar{\mathcal{X}}$ and $\mathcal{X}_w$ are not only unreliable but also have smaller variance than $\mathcal{X}''$. Furthermore, the manner in which the plotted points deviate from the diagonal suggests that $\bar{\mathcal{X}}$ and $\mathcal{X}_w$ are under-confident in both information scenarios.  The level of under-confidence 
is particularly startling in Figures \ref{RelEWANo} and \ref{RelOWANo} but 
decreases as information overlap is introduced in Figures \ref{RelEWAHigh} and \ref{RelOWAHigh}.  
  Given that averaging-like techniques do not behave like information aggregators, that is, they are not expanding, it is not surprising to see them perform better under high information overlap when aggregating information is less important for good performance. Table \ref{NoParams} shows the parameter estimates for $\mathcal{X}_w$ and $\mathcal{X}^*$.  The weights in $\mathcal{X}_w$ increase in the forecaster's amount of information and differ noticeably from the equal weights employed by $\bar{\mathcal{X}}$. 
More importantly, however, in both information scenarios $\alpha > 1$. This reflects the need to correct the under-confidence of $\mathcal{X}_w$. The resulting $\mathcal{X}^*$ is more reliable and confident as can be seen in Figures \ref{ELOPNoOverlap} and \ref{ELOPHighOverlap}. Furthermore, it behaves very similarly to the optimal aggregator $\mathcal{X}''$ under both information structures. 



In addition to performing visual assessment, the aggregators can be compared based on their out-of-sample average quadratic loss. To make this specific, let $\Y = (Y_1, \dots, Y_K)$ collect all the outcomes of the testing problems and $\boldsymbol{\mathcal{X}} = (\mathcal{X}_1, \dots, \mathcal{X}_K)$ be a vector of some aggregate forecasts for the same problems. Then, the average quadratic loss for this aggregator is
\begin{align*}
L\left(\Y, \boldsymbol{\mathcal{X}}\right) &= \frac{1}{K} \sum_{k=1}^K \left(Y_k - \mathcal{X}_k\right)^2.
\end{align*}
If the forecasts are probability estimates of binary outcomes, the above loss is known to have a decomposition that permits a closer analysis of  reliability and resolution \citep{Brier,murphy1973new}. The decomposition, however, is not limited to probability forecasts. To see this, suppose that the real-valued aggregate $\mathcal{X}_k \in \{f_1, \dots, f_I\}$ for some finite number $I$. Let $K_i$ be the number of times $f_i$ occurs, $\bar{Y}_i$ be the empirical average of $\{Y_k : \mathcal{X}_k = f_i\}$, and $\bar{Y} = \frac{1}{K} \sum_{k=1}^K Y_k$. Then,
\begin{align}
L\left(\Y, \boldsymbol{\mathcal{X}} \right)  &= \underbrace{\frac{1}{K} \sum_{i=1}^I K_i(f_i - \bar{Y}_i)^2}_\text{REL} - \underbrace{\frac{1}{K} \sum_{i=1}^I K_i(\bar{Y}_i-\bar{Y})^2}_\text{RES} + \underbrace{\frac{1}{K} \sum_{k=1}^K (Y_k - \bar{Y})^2}_\text{UNC}. \label{decomposition}
\end{align}
See the Appendix for the derivation of this decomposition. The three components of the decomposition are highly interpretable. In particular, low REL suggests high reliability. If the aggregate is reliable, then RES is approximately equal to the sample variance of the aggregate and 
is increasing in resolution.
The final term, UNC does not depend on the forecasts. This is the sample variance of $Y$ and therefore gives an approximate upper bound on the variance of any reliable forecast. As has been mentioned before, the goal is to maximize resolution subject to reliability. This decomposition shows how the quadratic loss addresses reliability and resolution simultaneously and therefore provides a convenient loss function for learning aggregation parameters.

\begin{table}[t!]
\centering
\caption{Synthetic Data. The average quadratic loss, $L(\Y,\boldsymbol{\mathcal{X}} )$ with its three additive components: reliability (REL), resolution (RES), and uncertainty (UNC). The final column, $s^2$ gives the estimated variance of the forecast.} 
\begin{tabular}{llrrrrr}
  \hline \hline
Scenario & Forecast & $L(\Y,\boldsymbol{\mathcal{X}} )$ & REL & RES & UNC & $s^2$ \\ 
  \hline
 \multirow{6}{*}{No Overlap} & Best Individual & 0.8024 & 0.0050 & 0.2108 & 1.0081 & 0.200\\ 
  & Median & 0.7322 & 0.2928 & 0.5688 & 1.0081& 0.046\\ 
  & $\bar{\mathcal{X}}$ & 0.7185 & 0.5140 & 0.8036 & 1.0081 & 0.032\\ 
  & $\mathcal{X}_w$ & 0.7016 & 0.2913 & 0.5979 & 1.0081 & 0.055\\ 
  & $\mathcal{X}^*$  & 0.1971 & 0.0022 & 0.8132 & 1.0081& 0.799\\ 
  & $\mathcal{X}''$ & 0.1969 & 0.0021 & 0.8132 & 1.0081 & 0.807\\ \rule{0pt}{2.9ex} 
  \multirow{6}{*}{High Overlap} &  Best Individual & 0.8141 & 0.0061 & 0.2195 & 1.0275 &0.199\\ 
  & Median & 0.8492 & 0.0087 & 0.1870 & 1.0275 &0.125 \\ 
  & $\bar{\mathcal{X}}$ & 0.8254 & 0.0137 & 0.2157 & 1.0275 &0.128\\ 
  & $\mathcal{X}_w$ & 0.7889 & 0.0166 & 0.2552 & 1.0275 &0.150 \\ 
  & $\mathcal{X}^*$  & 0.7758 & 0.0056 & 0.2573 & 1.0275 &0.228 \\ 
  &$\mathcal{X}''$ & 0.6837 & 0.0057 & 0.3496 & 1.0275 &0.318 \\ 
   \hline
\end{tabular}
\label{scenA}
\end{table}

Table \ref{scenA} presents the quadratic loss, its additive components, and the estimated variance $s^2$ for each of the different forecasts under both information scenarios. In addition to the aforementioned $\bar{\mathcal{X}}$, $\mathcal{X}_w$, $\mathcal{X}^*$, and $\mathcal{X}''$, the table also presents scores for the median forecast and the individual forecaster with the lowest quadratic loss. Even though the best individual is reliable by construction, it is highly unresolute and hence gains an overall poor quadratic loss. 
%
 Under no information overlap, however, this individual is better than both the median and $\bar{\mathcal{X}}$ because these aggregators assign too much importance to the individual forecasters with very little information.  As predicted by Theorem \ref{contraction},  the median and the averaging aggregators $\bar{\mathcal{X}}$ and $\mathcal{X}_w$ are neither reliable nor expanding. The remaining two aggregators, namely $\mathcal{X}^*$ and $\mathcal{X}''$, on the other hand, are reliable and expanding. Table \ref{NoParams} shows that $\mathcal{X}^*$ is in fact  almost equivalent to $\mathcal{X}''$ under no information overlap. Under high information overlap, however, $\mathcal{X}''$ gains slight advantage over $\mathcal{X}^*$. In this case $\mathcal{X}^*$ cannot take the same form as $\mathcal{X}''$. Consequently, it has an estimated variance of $0.228$ which is well below the  amount of information known to the group, namely $0.320$. It fails to use information optimally because it cannot subtract off the shared information $X_1$ and hence avoid double-counting of information. However, despite it using information less efficiently, it is as reliable as $\mathcal{X}''$. 

Of course, under the Gaussian model, $\mathcal{X}^*$ may seem redundant because the optimal  $\mathcal{X}''$ can be computed directly. In practice, however, $\bSigma$ is not known and must be estimated under a non-trivial semidefinite constraint (see \citealt{satopaamodeling2} for more details). Given that this involves a total of $\binom{N}{2} + N$ parameters, the estimation task is challenging even for moderately large $N$, say, greater than $100$. Furthermore, accurately estimating such a large number of parameters requires the forecasters to attend a large number of prediction problems. Applying $\mathcal{X}^*$ instead is significantly easier because it involves only $N+1$ parameters that can be estimated via a standard quadratic program (\ref{secondProblem}).  Therefore this aggregator scales better to large groups of forecasters. On the other hand, problem (\ref{secondProblem}) requires a training set with known outcomes whereas $\bSigma$ can be learned from the forecasts alone. Therefore the two aggregators serve somewhat different purposes and should be considered complementary rather than competitive.

\section{CASE STUDY: CONCRETE COMPRESSIVE STRENGTH} \label{application}

Concrete is the most important material in civil engineering. One of its key properties is compressive strength that depends on the water-to-cement ratio but also on several other ingredients. \cite{yeh1998modeling} illustrated this by statistically predicting compressive strength based on age and seven mixture ingredients. The associated dataset is freely available at the UC Irvine Machine Learning Repository \citep{Lichman:2013} and consists of $1,030$ observations with the following information:

\begin{equation}
\begin{array}{ll r lr}
&&Y: & \text{Compressive Strength}&\\
\ldelim\{{8}{2.5em}[$\mathcal{M}_F$] &\ldelim\{{4}{2.5em}[$\mathcal{M}_1$] &v_1: &  \text{Cement (kg in a $m^3$ mixture)}&  \\
&&v_2: & \text{Coarse Aggregate (kg in a $m^3$ mixture)}&\\
  &&v_3: &  \text{Fly Ash (kg in a $m^3$ mixture)}&  \rdelim\}{4}{2em}[$\mathcal{M}_3$] \\
&&v_4: & \text{Water (kg in a $m^3$ mixture)}&\\
&\ldelim\{{4}{2.5em}[$\mathcal{M}_2$] &v_5: &  \text{Superplasticizer (kg in a $m^3$ mixture)}&  \\
&&v_6: & \text{Fine Aggregate  (kg in a $m^3$ mixture)}&\\
&&v_7: & \text{Blast Furnace Slag (kg in a $m^3$ mixture)}&\\
&&v_8: & \text{Age (days)}&
\end{array}\label{realData}
\end{equation}
\noindent
This particular dataset is appropriate for illustrating our results because it is simple yet large enough to allow the computation of reliability diagrams and the individual components of the average quadratic loss.

The individual forecasters are emulated with three linear regression models, $\mathcal{M}_1$, $\mathcal{M}_2$, and $\mathcal{M}_3$, that predict $Y$ based on different sets of predictors. In particular, model $\mathcal{M}_1$ only uses predictors $v_1, v_2, v_3, v_4$, whereas model $\mathcal{M}_2$ uses the remaining predictors $v_5, v_6, v_7, v_8$. Therefore their predictor sets are non-overlapping. The third model $\mathcal{M}_3$ uses the middle four predictors $v_3, v_4, v_5, v_6$, and hence has significant overlap with the other two models. The results are compared against a linear regression model $\mathcal{M}_F$ that has access to all eight predictors. This is not an aggregator and only represents the extent to which the predictors can explain the outcome $Y$. Therefore it provides interpretation and scale. The predictor sets corresponding to the different models are summarized by the curly braces in (\ref{realData}). Overall, this setup can be viewed as a real-valued equivalent of the case study in \cite{Ranjan08} who aggregate probability forecasts from three different logistic regression models. 

The evaluation is based on a $10$-fold cross validation. The models $\mathcal{M}_1$, $\mathcal{M}_2$, and $\mathcal{M}_3$ are first trained on one half of the training set and then used to make  predictions for the second half and the entire testing set. Next, the aggregators are  trained on the models' predictions over the second half of the training set. Finally, the trained aggregators are tested on the models' predictions over the testing set. Therefore all the following results, apart from the parameter estimates, represent out-of-sample performance. Similarly to Section \ref{simulation}, the evaluation is performed separately under two different information structures: the \textit{No Information Overlap} scenario considers only predictions from models $\mathcal{M}_1$ and $\mathcal{M}_2$, whereas the \textit{High Information Overlap} scenario involves only predictions from models $\mathcal{M}_1$ and $\mathcal{M}_3$.

\begin{figure}[t]
        \centering
        \begin{subfigure}[b]{0.241\textwidth}
                \includegraphics[width=\textwidth]{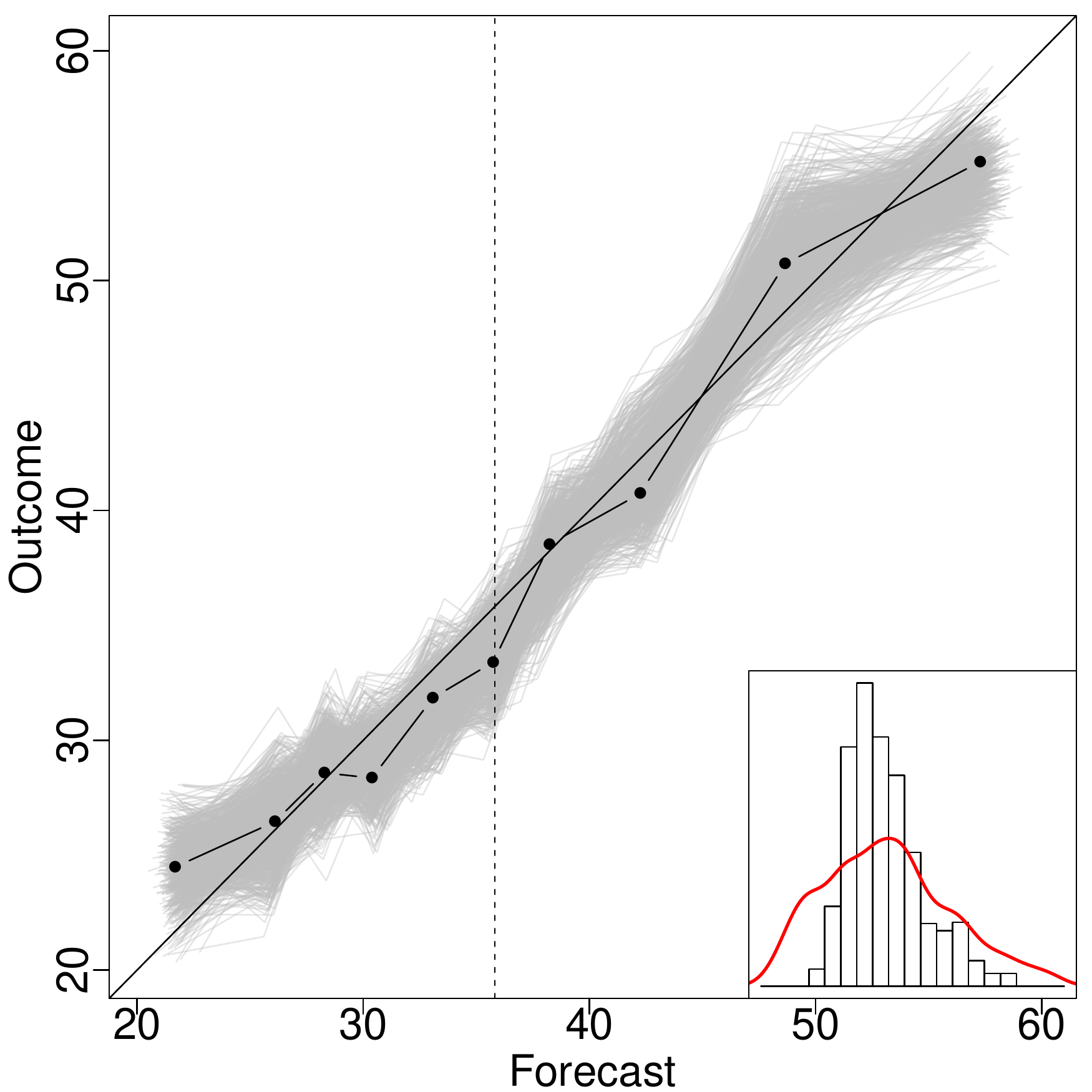}
                \caption{$\mathcal{M}_1$}
                \label{fig:gull}
        \end{subfigure}%
        ~ 
        \begin{subfigure}[b]{0.241\textwidth}
                \includegraphics[width=\textwidth]{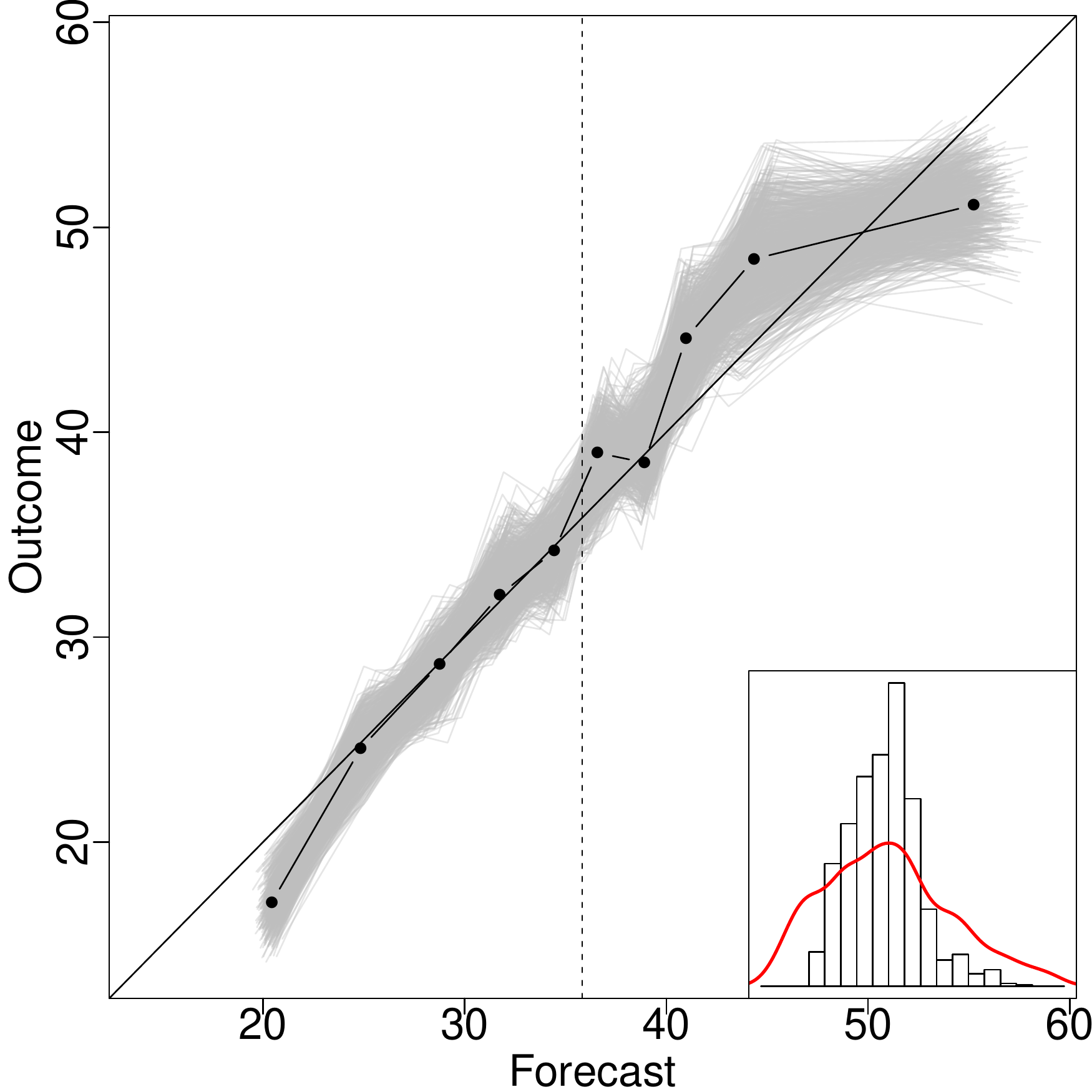}
                \caption{$\mathcal{M}_2$}
                \label{fig:tiger}
        \end{subfigure}
        \begin{subfigure}[b]{0.241\textwidth}
                \includegraphics[width=\textwidth]{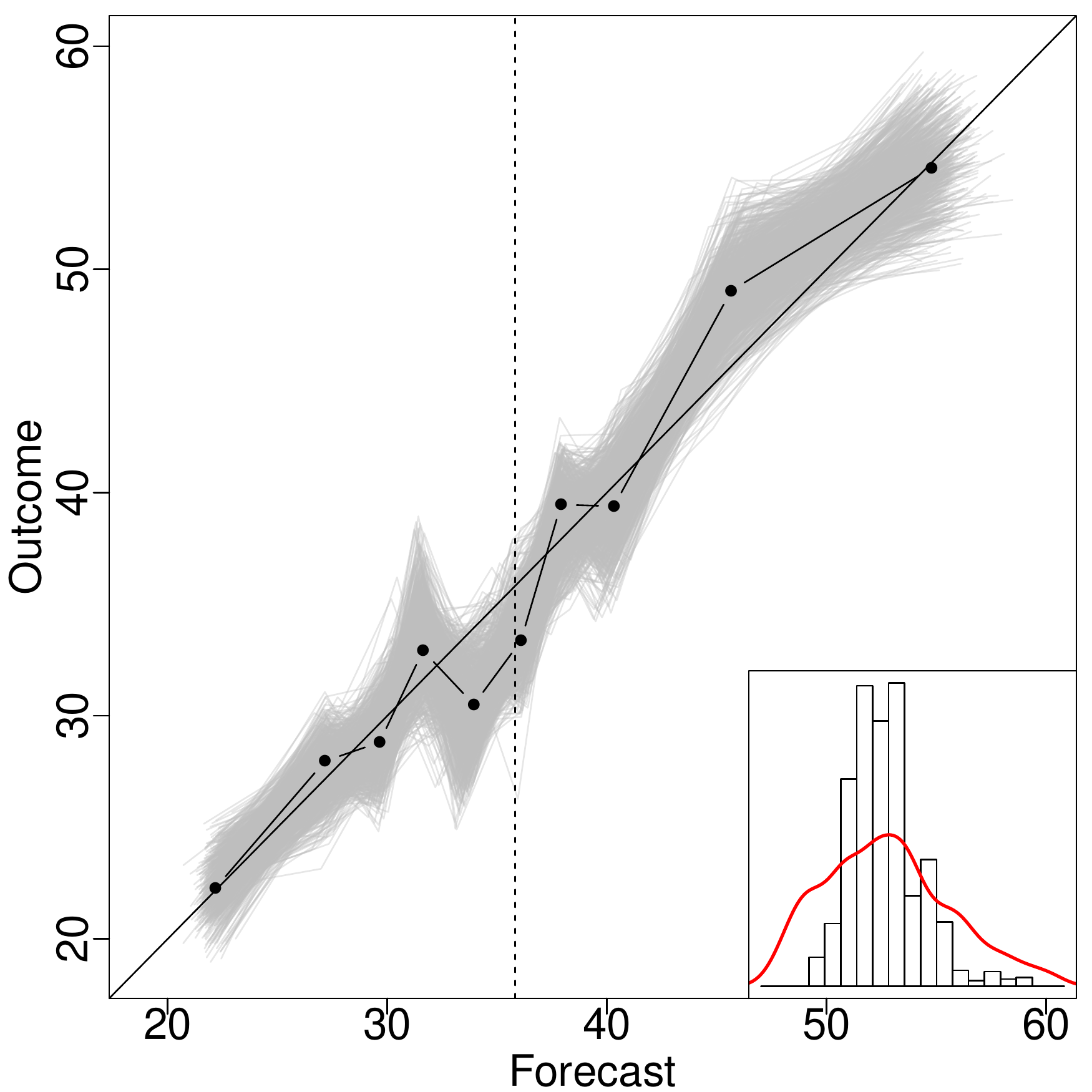}
                \caption{$\mathcal{M}_3$}
                \label{fig:tiger}
        \end{subfigure}
                \begin{subfigure}[b]{0.241\textwidth}
                \includegraphics[width=\textwidth]{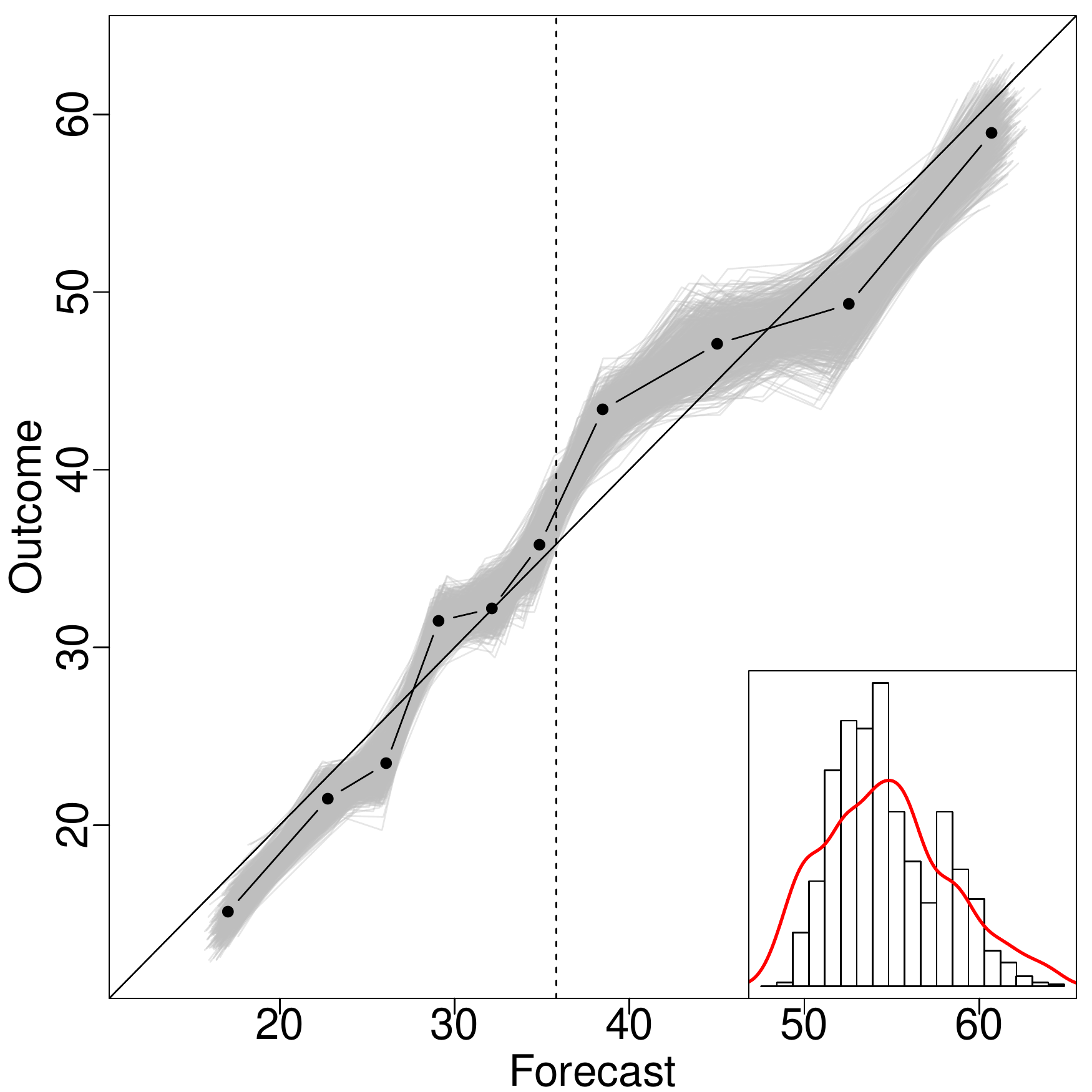}
                \caption{$\mathcal{M}_F$}
                \label{RelDiagramNoF}
        \end{subfigure}           
          \caption{Real-World Data. Out-of-sample reliability of the individual models.}
                \label{RelDiagramMo}
\end{figure}
\begin{figure}[t]
        \centering
 
        \begin{subfigure}[b]{0.323\textwidth}
                \includegraphics[width=\textwidth]{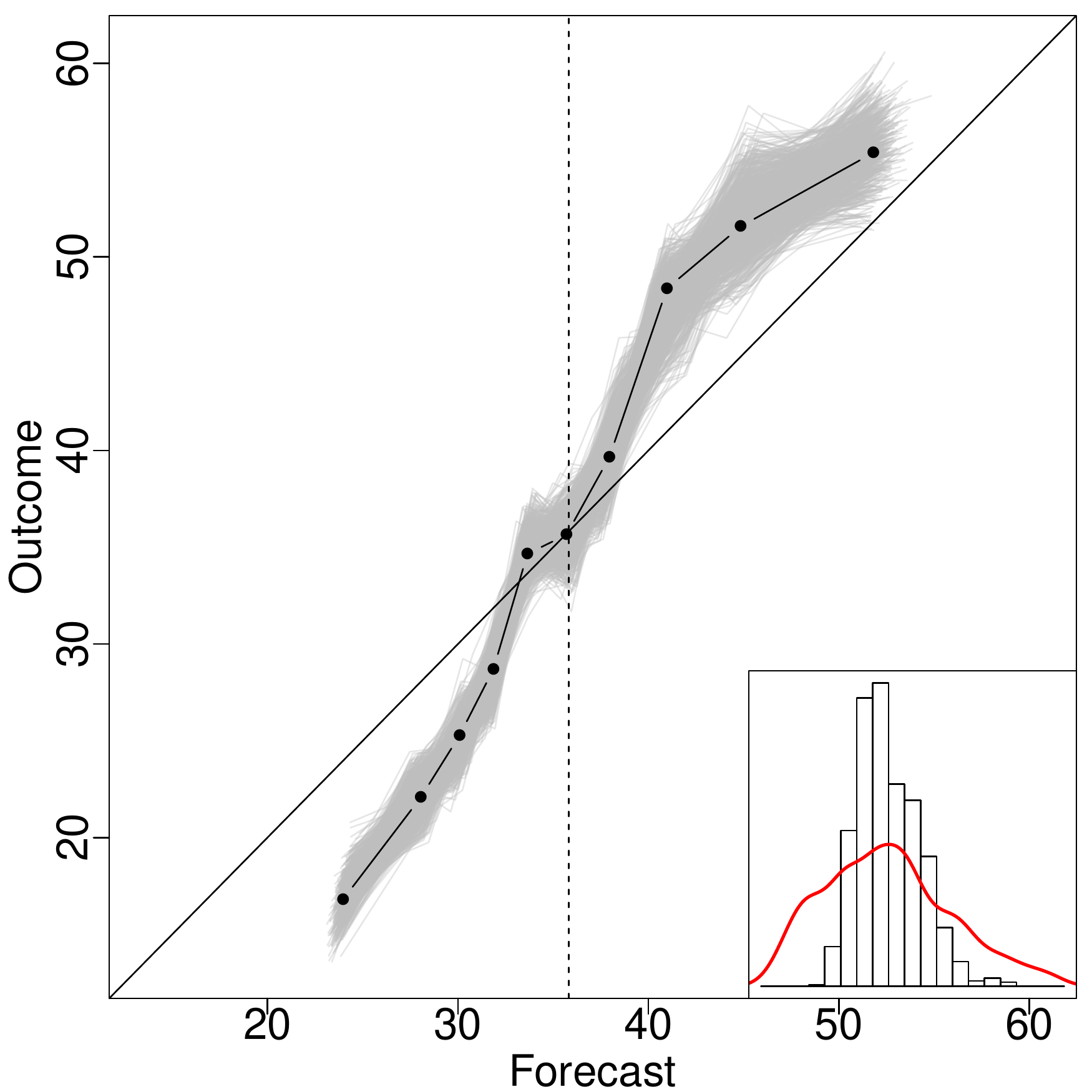}
                \caption{$\bar{\mathcal{X}}$}
                \label{fig:mouse}
        \end{subfigure}
                  \begin{subfigure}[b]{0.323\textwidth}
                \includegraphics[width=\textwidth]{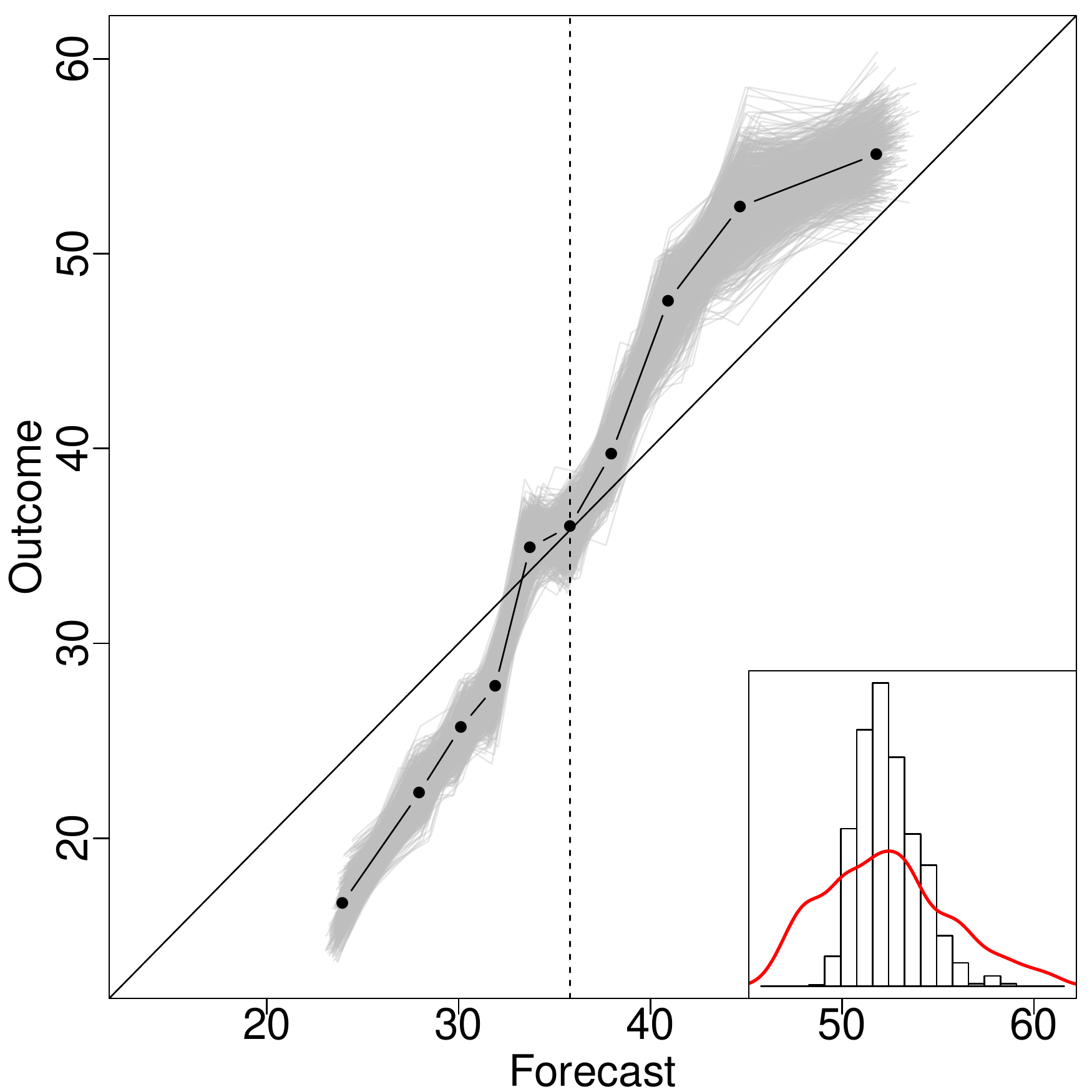}
                \caption{$\mathcal{X}_w$}
                \label{fig:gull}
        \end{subfigure}%
         \begin{subfigure}[b]{0.323\textwidth}
                \includegraphics[width=\textwidth]{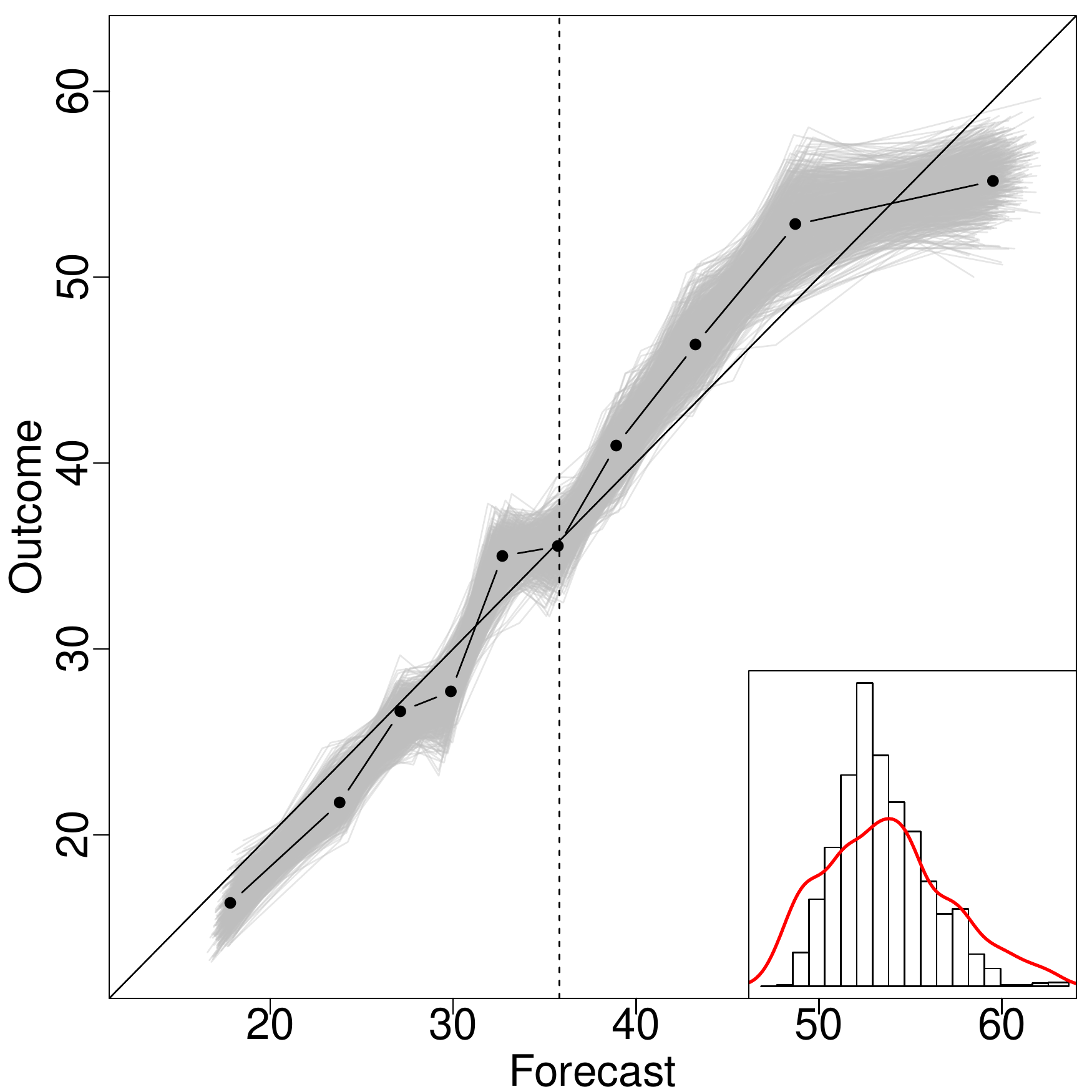}
                \caption{$\mathcal{X}^*$}
                \label{RelDiagramNoE}
             \end{subfigure}
          \caption{Real-World Data. Out-of-sample reliability of aggregators under no information overlap. }
           \label{RelDiagramNo}
\end{figure}
\begin{figure}[t]
        \centering
        \begin{subfigure}[b]{0.323\textwidth}
                \includegraphics[width=\textwidth]{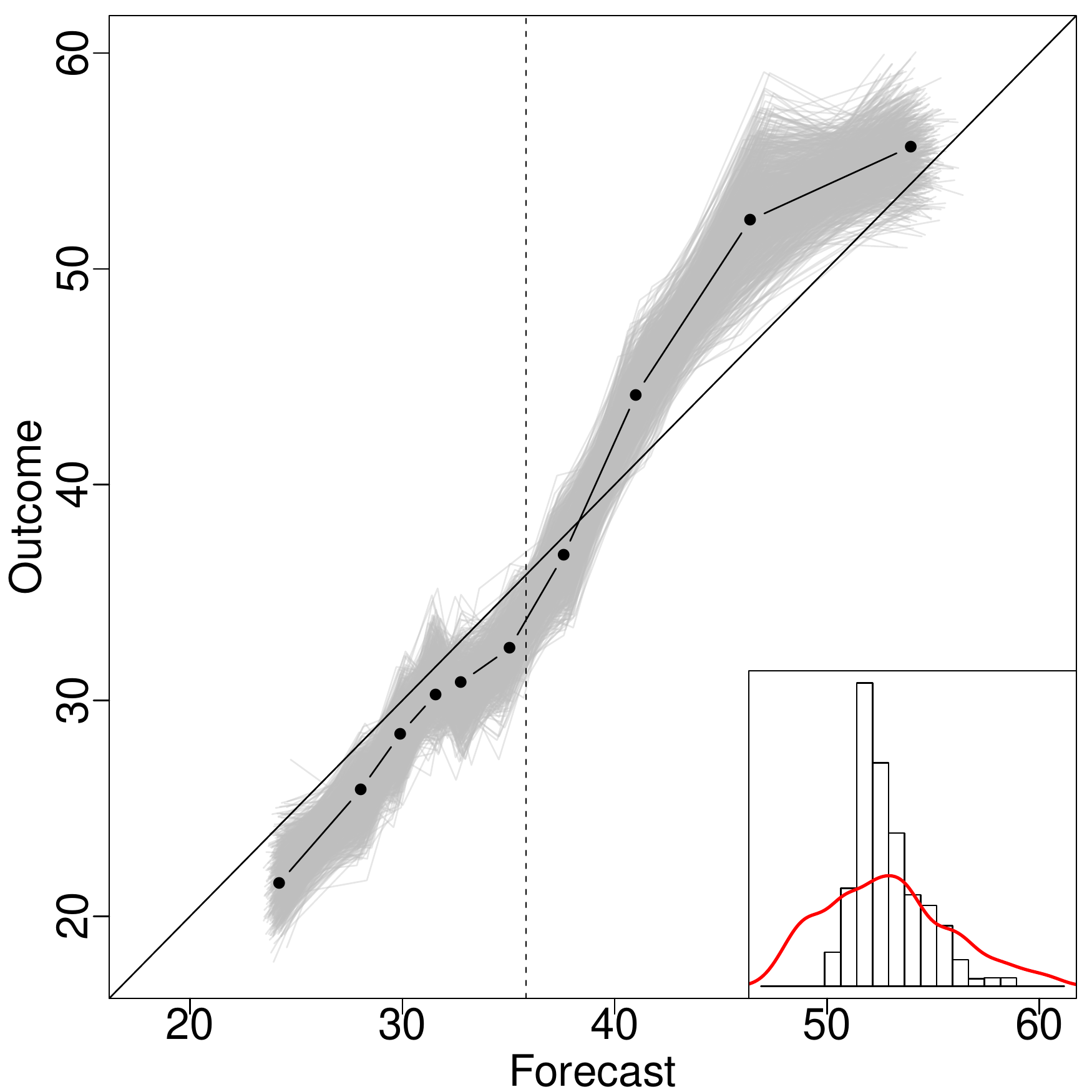}
                \caption{$\bar{\mathcal{X}}$}
                \label{fig:mouse}
        \end{subfigure}     
                \begin{subfigure}[b]{0.323\textwidth}
                \includegraphics[width=\textwidth]{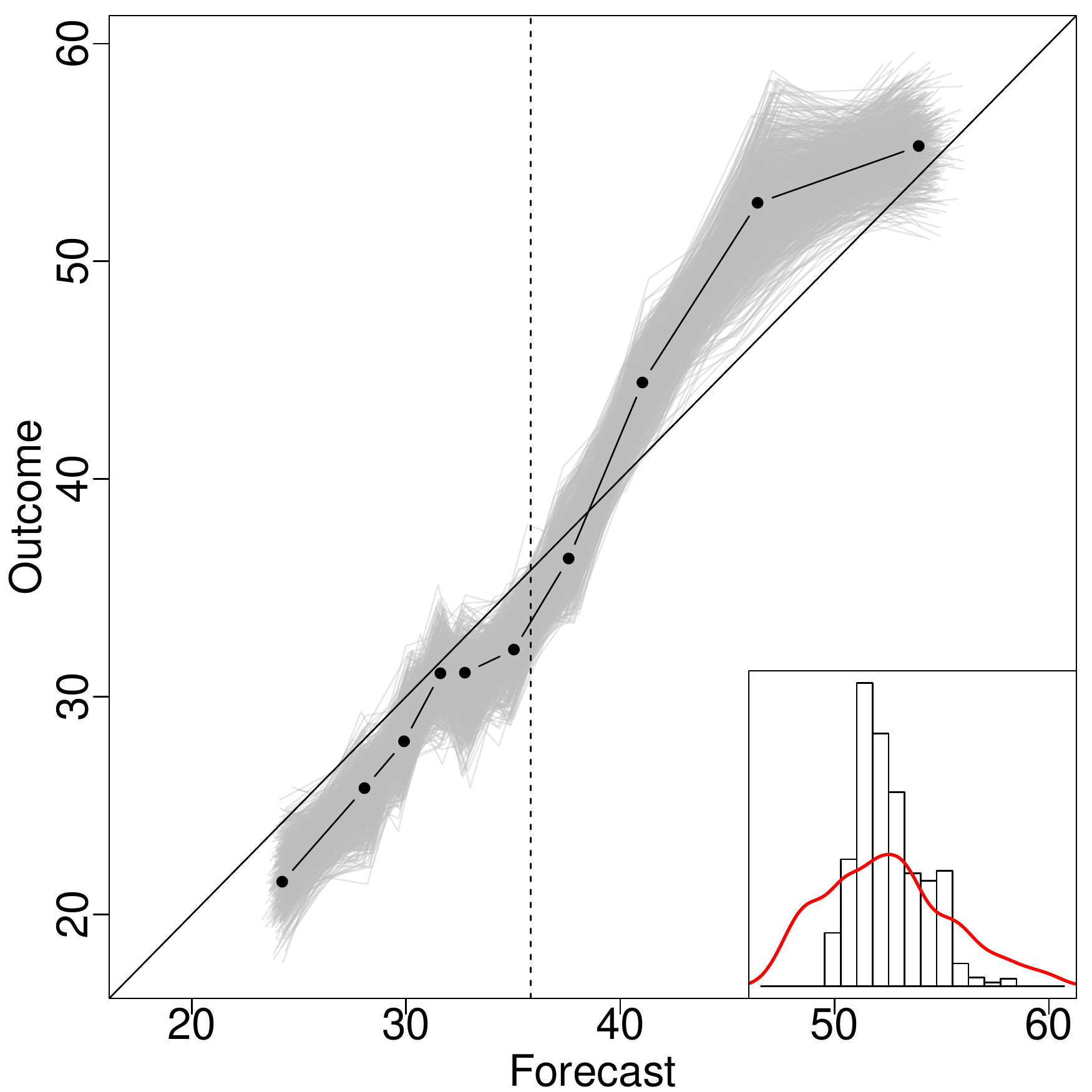}
                \caption{$\mathcal{X}_w$}
                \label{fig:gull}
        \end{subfigure}%
        \begin{subfigure}[b]{0.323\textwidth}
                \includegraphics[width=\textwidth]{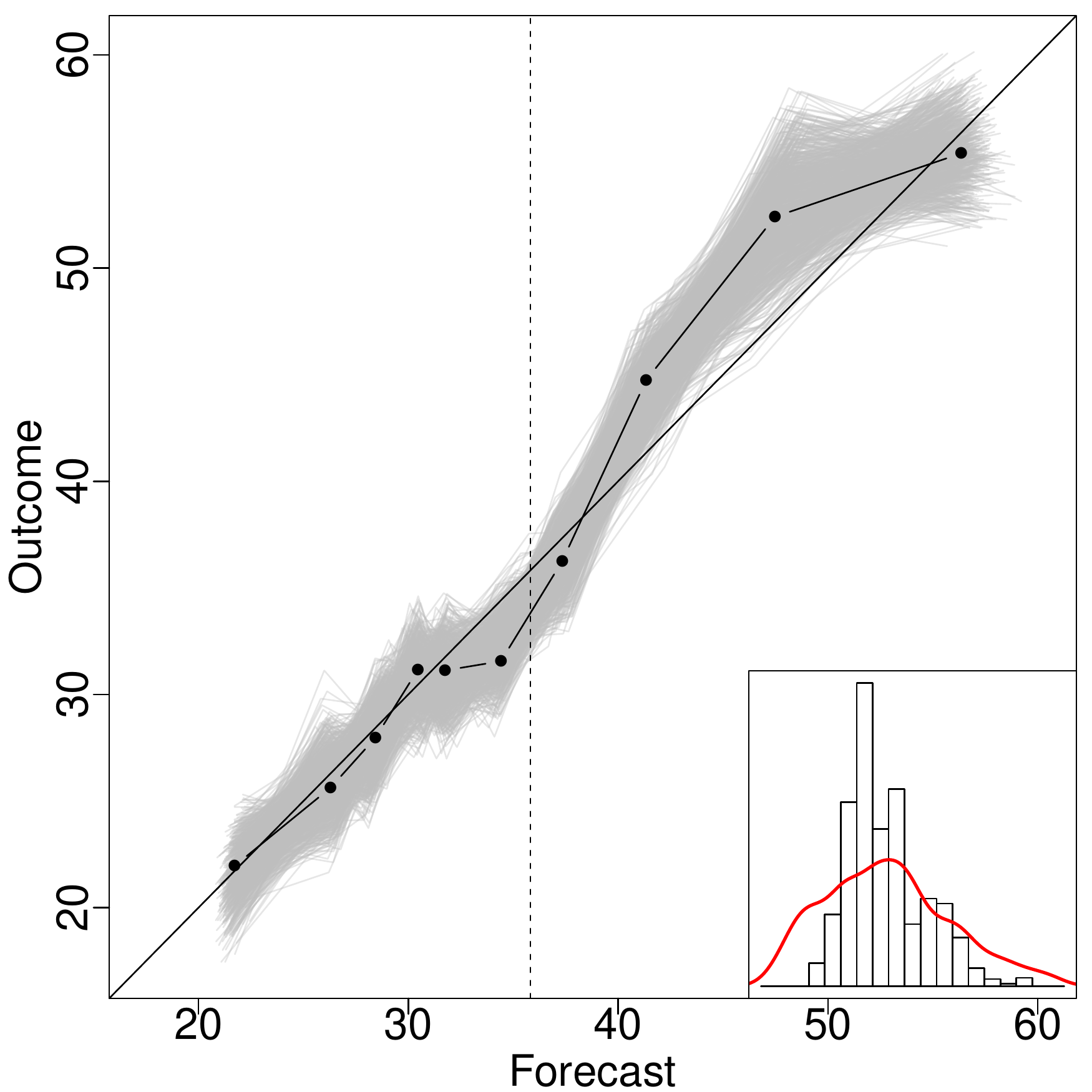}
                \caption{$\mathcal{X}^*$}
                \label{DepEOLPConrete}
        \end{subfigure}
          \caption{Real-World Data. Out-of-sample reliability of aggregators under high information overlap. }
               \label{RelDiagramHigh}
\end{figure}

Figures \ref{RelDiagramMo}, \ref{RelDiagramNo}, and \ref{RelDiagramHigh} present the reliability diagrams of the individual models and the aggregators under no and high information overlap, respectively. Unlike in Section \ref{simulation}, the marginal distribution of $Y$ is not known. Therefore the red curve over the inlined histogram represents the empirical distribution of $Y$. Similarly, the dashed vertical line represents the sample average of the outcomes instead of the marginal mean $\mu_0$. According to these plots, the individual forecasts are mostly reliable, except at extremely small or large forecasts. The averaging aggregators $\bar{\mathcal{X}}$ and $\mathcal{X}_w$, on the other hand, are both unreliable and under-confident. Similarly to Section \ref{simulation} and in accordance with Theorem \ref{contraction}, this under-confidence decreases as the forecasters' information overlap increases from Figure  \ref{RelDiagramNo} to Figure \ref{RelDiagramHigh}.  Table \ref{NoParamsReal} gives the parameter estimates for $\mathcal{X}_w$ and $\mathcal{X}^*$. These aggregators employ very similar weights. In both information scenarios $\alpha > 1$, suggesting that $\mathcal{X}_w$ is under-confident and should be extremized as it is. Based on Figures \ref{RelDiagramNoE} and  \ref{DepEOLPConrete}, the resulting aggregator $\mathcal{X}^*$ is noticeably more reliable and appears to approximate the empirical distribution of $Y$ quite closely. Simply based on visual assessment $\mathcal{X}^*$ performs as well as $\mathcal{M}_F$ under low information overlap but  loses some resolution once  overlap is introduced. This makes sense because the models considered in the high information overlap scenario, namely $\mathcal{M}_1$ and $\mathcal{M}_3$ have access only to the first six predictors while $\mathcal{M}_F$ uses all eight predictors and hence should have a higher level of information.

\begin{table}[t!]
\centering
\caption{Real-World Data. Estimated parameter values.} 
\begin{tabular}{llrrrr}
   \hline \hline
Scenario & Forecast & $\mu_0$ & $\alpha$ & $w_1$ & $w_2$\\
  \hline
\multirow{2}{*}{No Overlap} & $\mathcal{X}_w$ &  &  & 0.5327 & 0.4673 \\ 
&  $\mathcal{X}^*$ & -36.2051 & 1.6950 & 0.5269 & 0.4731 \\  \rule{0pt}{2.9ex} 
\hspace{-0.2em}\multirow{2}{*}{High Overlap}  & $\mathcal{X}_w$ &  &  & 0.5931 & 0.4069 \\ 
 & $\mathcal{X}^*$ & -37.6776 & 1.4382 & 0.5375 & 0.4625 \\ 
   \hline
\end{tabular}
\label{NoParamsReal}
\end{table}

\begin{table}[t!]
\centering
\caption{Real-World Data. The average quadratic loss, $L(\Y,\boldsymbol{\mathcal{X}} )$ with its three additive components: reliability (REL), resolution (RES), and uncertainty (UNC). The final column, $s^2$ gives the estimated variance of the forecast. } 
\begin{tabular}{llrrrrr}
  \hline \hline
Scenario &  Forecast & $L(\Y,\boldsymbol{\mathcal{X}} )$ & REL & RES & UNC & $s^2$\\ 
  \hline
 &  $\mathcal{M}_1$ & 187.80 & 9.70 & 100.72 & 278.81 & 82.83 \\ 
& $\mathcal{M}_2$  & 185.74 & 12.01 & 105.08 & 278.81 & 92.51 \\ 
  & $\mathcal{M}_3$ & 197.03 & 12.81 & 94.59 & 278.81 & 73.27 \\ 
&$\mathcal{M}_F$  & 110.91 & 9.46 & 177.36 & 278.81 & 157.87 \\ \rule{0pt}{2.9ex} 
\multirow{3}{*}{No Overlap} &  $\bar{\mathcal{X}}$ & 155.69 & 30.99 & 154.10 & 278.81 & 56.33 \\ 
 & $\mathcal{X}_w$ & 156.32 & 31.45 & 153.94 & 278.81 & 56.21 \\ 
  &$\mathcal{X}^*$ & 133.23 & 9.86 & 155.45 & 278.81 & 161.89 \\ \rule{0pt}{2.9ex} 
 \multirow{3}{*}{High Overlap}  & $\bar{\mathcal{X}}$ & 177.45 & 16.77 & 118.13 & 278.81 & 61.92 \\ 
  & $\mathcal{X}_w$ & 176.59 & 14.37 & 116.59 & 278.81 & 63.32 \\ 
 & $\mathcal{X}^*$ & 169.92 & 8.20 & 117.09 & 278.81 & 128.69 \\ 
\hline
\end{tabular}
\label{NoTbl}
\end{table}

%
Table \ref{NoTbl} provides a numerical comparison by presenting the average quadratic loss, its additive components, and the estimated variance $s^2$ for the individual models and the competing aggregators.  Given that all aggregators perform better than the individual forecasters, aggregation is generally beneficial. However, there are large performance differences among the aggregators. In particular, the variances of $\bar{\mathcal{X}}$ and $\mathcal{X}_w$  do not exceed that of the individual forecasters', suggesting that neither of them is expanding. Furthermore, they are much less reliable than the individual forecasters. In contrast, $\mathcal{X}^*$ is able to maintain the forecasters' level of reliability. Even though this aggregator is expanding, it is less resolute and has a lower variance than $\mathcal{M}_F$ under high information overlap. This can be expected because in the high information overlap scenario $\mathcal{X}^*$ has access only to a subset of the information that $\mathcal{M}_F$ uses. Under no information overlap, all the predictors are  used by the individual forecasters, but this does not mean that this information is actually revealed to $\mathcal{X}^*$ through the reported forecasts. 

\section{SUMMARY AND DISCUSSION} \label{conclusion}

This paper discussed forecast aggregation under a general probability model, called the partial information framework. The forecasts and outcomes were assumed to have a joint distribution but no restrictions were placed on their dependence structure. The analysis led to an enumeration (Theorem \ref{optimal}) of several properties of optimal aggregation. Even though the optimal aggregator is typically intractable in practice, its properties provide guidance for developing and understanding other aggregators that are more feasible in practice. In this paper these properties shed light on the class of weighted averages of any type of univariate forecasts. Even though these averages are marginally consistent, they fail to satisfy two of the optimality properties, namely reliability and variance expansion (Theorem \ref{contraction}). As a result, they are under-confident in a sense that they are overly close to the marginal mean. This shortcoming can be naturally alleviated by extremizing, that is, by shifting the weighted average further away from the marginal mean.  Section \ref{extremization} introduced a simple linear procedure (Equation \ref{firstProblem}) that extremizes the weighted average of real-valued forecasts and maintains marginal consistency. This procedure and the theoretical results were illustrated on synthetic (Section \ref{simulation}) and real-world data (Section \ref{application}). In both cases the optimally weighted average was shown to be both unreliable and under-confident, especially when the forecasters used very different sets of information. Fortunately, extremization was able to largely correct these drawbacks and provide transformed aggregates that were both reliable and more resolute. 


Forecast aggregation literature by and large agrees that the goal is to collect and combine  information from different forecasters (see, e.g., \citealt{dawid1995coherent, armstrong2, forlines2012heuristics}). At the same time aggregation continues to be performed via weighted averaging or perhaps some other measure of central tendency, such as the median \citep{levins1966strategy, armstrong2, lobo2010human}. Section \ref{contraction} explained that these popular techniques do not behave like aggregators of information. Instead, they are designed to reduce measurement error which is philosophically very different from information diversity \citep{satopaamodeling2}. Therefore some details of their workings seem to have been misunderstood. Unfortunately, it is unlikely that this paper will prevent aggregation with measures of central tendency all together. However, it is hoped that our contributions will at least prompt interest and provide direction in discovering alternative aggregation techniques. 

This paper illustrated that good information aggregation can arise from a simple linear transformation that extremizes the weighted average. Of course, under a large number of prediction problems, a non-linear extremizing function can lead to further improvements in aggregation. The linear function, however, is a simple and natural starting point that suffices for illustrating the benefits of extremizing. Is extremizing then guaranteed to be beneficial in every prediction task? Probably not. 
Therefore, for the sake of applications, it is important to discuss conditions under which extremizing is likely to improve the commonly used aggregators. Item \ref{underconfB}) of Theorem \ref{contraction} and the empirical results in Sections \ref{simulation} and \ref{application} suggest that extremizing is likely to be more beneficial under no or low information overlap. This aligns with  \cite{satopaamodeling} who use the Gaussian partial information model to show empirically that extremizing probability forecasts becomes more important a) as the amount of the forecasters' combined information increases, and b) as the forecasters' information sets become more diverse. This means that, for instance, the average forecast of team members working in close collaboration require little extremizing whereas forecasts coming from widely different sources must be heavily extremized.

Unfortunately, the amount and direction of extremization depends on a training set with known outcomes. Such a training set may not always be available. In the most extreme case  the decision-maker may have only a set of forecasts of a single unknown outcome. How should the forecasts be aggregated in such a low-data setting? The results in this paper suggest that any type of weighted average (or some other measure of central tendency) is a poor choice. A better alternative was discussed by \cite{satopaamodeling}. They assume that the forecasters' covariance matrix is compound symmetric and then aggregate the probability forecasts with the optimal aggregator under the corresponding Gaussian partial information model. 
 Developing more general aggregators that place less constraints on the joint dependence structure while satisfying at least two of the optimality properties of Theorem \ref{optimal} is certainly an interesting future research direction. The first step is to develop a simple aggregator that is both marginally consistent and expanding. Finding an aggregator that maintains forecasters' reliability seems more difficult. 



%
%
%
%
%
%
%
%
%
%
%

\appendix
\section{APPENDIX}
\subsection{Proof of Theorem \ref{optimal}}
\begin{enumerate}[i)]
\item The law of total expectation gives:
\begin{align*}
\E(\mathcal{X}'') &= \E[\E(Y|\mathcal{X}'')] =  \E(Y) = \mu_0.
\end{align*}

\item Recall that $ \mathcal{X}'' = \E(Y | \F'')$, $\mathcal{X}'' \in \F''$, and $\F'' = \sigma(X_1, \dots, X_N)$. Then,
\begin{align*}
&\hspace{1.3em}  \E(Y | \mathcal{X}'') \\
 &= \E[\E(Y|\mathcal{X}'',\F'')|\mathcal{X}''] & \text{ (as $\mathcal{X}'' \in \F''$)}\\
&= \E[\E(Y|\F'')|\mathcal{X}'']\\
&= \E(\mathcal{X}''|\mathcal{X}'') \\
&= \mathcal{X}''.
\end{align*}

\item This relies on the observation that $\sigma(X_m) = \F_m \subseteq \F'' = \sigma(X_1, \dots, X_N)$. Then,
\begin{align*}
\delta_{max} &=\Var(X_m)\\
 &= \E\left(X_m^2\right) - \mu_0^2\\
 &= \E[\E(Y|\F_m)X_m] - \mu_0^2 & \text{(as $X_m = \E(Y|\F_m)$)}\\
 &= \E\{\E[\E(Y|\F'')|\F_m]X_m\} - \mu_0^2 & \text{(the smallest $\sigma$-field wins)}\\
 &= \E[\E(\mathcal{X}''|\F_m)X_m] - \mu_0^2 \\
 &= \E[\E(\mathcal{X}''X_m|\F_m)] - \mu_0^2 \\
 &= \E(\mathcal{X}''X_m) - \mu_0^2 & \text{(reverse iterated expectation)} \\
 &= \E[(\mathcal{X}''-\mu_0)(X_m-\mu_0)]  \\
 &\leq \sqrt{\Var(\mathcal{X}'')\delta_{max} } & \text{(by the Cauchy-Schwarz inequality).} 
\end{align*}
Squaring and diving both sides by $\delta_{max} $ gives the desired result. 

\end{enumerate}
\qed

\subsection{Proof of Theorem \ref{contraction}}
Items ii) and iii) are generalizations of the proof in \cite{Ranjan08}. 
\begin{enumerate}[i)]
\item
This follows from direct computation:
\begin{align*}
\E(\mathcal{X}_w) = \E(\w' \X) &= \w' \E(\X) = \mu_0 \w' \one_N = \mu_0.
\end{align*}

\item Consider some reliable aggregate $\mathcal{X}$ such that $\E(Y | \mathcal{X}) = \mathcal{X}$. Then,
\begin{align*}
&\hspace{1.3em} \E[(Y-\mathcal{X})^2]\\
 &= \E\left\{\E\left[(Y-\mathcal{X})^2|\mathcal{X}\right]\right\}\\
&= \E\left[\E\left(Y^2-2Y\mathcal{X}+\mathcal{X}^2|\mathcal{X} \right)\right]\\
&= \E\left[\E\left(Y^2|\mathcal{X}\right)-\mathcal{X}^2\right]\\
&= \E(Y^2)-\E(\mathcal{X}^2).
\end{align*}
The rest of the proof shows that if $\mathcal{X} = \mathcal{X}_w = \w'\X$, then the above identity cannot hold. This gives a contradiction and hence proves the desired result. First, note that $\sum_{i=1}^N\sum_{j=1}^Nw_iw_j = 1$. Then,
\begin{align*}
&\hspace{1.3em}  \E\left[\left(Y-\mathcal{X}_w\right)^2\right]\\
 &= \E\left[\left(Y-\w'\X\right)^2\right] \\
 &= \E\left\{\left[\sum_{j=1}^Nw_j (Y-X_j)\right]^2\right\} \\
&= \sum_{i=1}^N\sum_{j=1}^Nw_iw_j \E[(Y-X_i)(Y-X_j)] \\
&= \sum_{i=1}^N\sum_{j=1}^Nw_iw_j \E\left(Y^2-YX_i-YX_j+X_jX_i\right) \\
&= \sum_{i=1}^N\sum_{j=1}^Nw_iw_j \E\left[ \E\left( Y^2|X_i \right)-\E\left( YX_i|X_i \right)-\E\left( YX_j|X_j\right)+X_jX_i \right] \\
&= \sum_{i=1}^N\sum_{j=1}^Nw_iw_j \E\left[ \E\left( Y^2|X_i \right)-X_i^2-X_j^2+X_jX_i \right] \\
&= \sum_{i=1}^N\sum_{j=1}^Nw_iw_j \E\left[\E\left(Y^2|X_i\right) +\left(X_jX_i-X_jX_i\right)-X_i^2-X_j^2+X_jX_i\right] \\
&= \sum_{i=1}^N\sum_{j=1}^Nw_iw_j \E\left[\E\left(Y^2|X_i\right) -X_jX_i-\left(X_i-X_j\right)^2\right] \\
&= \sum_{i=1}^N\sum_{j=1}^Nw_iw_j \E\left[\E\left(Y^2|X_i\right) -X_jX_i\right]-\sum_{i=1}^N\sum_{j=1}^Nw_iw_j\E\left[\left(X_i-X_j\right)^2\right] \\
&=  \E\left(Y^2\right) - \sum_{i=1}^N\sum_{j=1}^Nw_iw_j \E\left(X_jX_i\right)-\sum_{i=1}^N\sum_{j=1}^Nw_iw_j\E\left[\left(X_i-X_j\right)^2\right] \\
&=  \E\left(Y^2\right) -  \E\left(\w'\X \X'\w\right)-\sum_{i=1}^N\sum_{j=1}^Nw_iw_j\E\left[\left(X_i-X_j\right)^2\right] \\
&= \left[\E\left(Y^2\right) -  \E\left(\mathcal{X}_w^2\right) \right]-\sum_{i=1}^N\sum_{j=1}^Nw_iw_j\E\left[\left(X_i-X_j\right)^2\right].
\end{align*}
This leads to a contradiction because the double sum on the final line is strictly positive as long as there exists a forecast pair $i \neq j$ such that $\P(X_i \neq X_j) > 0$ and $w_i, w_j > 0$. 

\item The fact that $\E(\mathcal{X}_w') = \mu_0$ follows similarly to the proof of item i) of Theorem \ref{optimal}. This item continues under the conditions of the previous item. Therefore it can be assumed that $\mathcal{X}_w$ is not calibrated, that is, $\P(\mathcal{X}_w' \neq \mathcal{X}_w) > 0$. Then,
\begin{align*}
&\hspace{1.3em}   \E \left[ \left( Y - \mathcal{X}_w\right)^2\right]\\
 &= \E \left( Y^2 - 2Y\mathcal{X}_w + \mathcal{X}_w^2\right)\\
&= \E \left( Y^2 + 2\left(\mathcal{X}_w'^2-\mathcal{X}_w'^2\right)- 2Y\mathcal{X}_w + \mathcal{X}_w^2\right)\\
&= \E \left( Y^2 -2Y\mathcal{X}_w'+2\mathcal{X}_w'^2- 2\mathcal{X}_w'\mathcal{X}_w + \mathcal{X}_w^2\right)\\
&= \E\left[\left(Y-\mathcal{X}_w'\right)^2\right] + \E\left[\left(\mathcal{X}_w-\mathcal{X}_w'\right)^2\right] \\
&= \E(Y^2)-\E(\mathcal{X}_w'^2) + \E\left[\left(\mathcal{X}_w-\mathcal{X}_w'\right)^2\right] & \text{(because $\mathcal{X}_w'$ is reliable)}\\
&> \E(Y^2)-\E(\mathcal{X}_w'^2).
\end{align*}
Furthermore, from the previous item, $\E \left[ \left( Y - \mathcal{X}_w\right)^2\right] < \E(Y^2)-\E(\mathcal{X}_w^2)$. Putting this all together gives
\begin{align*}
&&\E(Y^2)-\E(\mathcal{X}_w'^2)  &< \E(Y^2)-\E(\mathcal{X}_w^2)\\
\Leftrightarrow && \E(\mathcal{X}_w'^2) - \mu_0^2  &> \E(\mathcal{X}_w^2) - \mu_0^2\\
\Leftrightarrow && \Var(\mathcal{X}_w')  &> \Var(\mathcal{X}_w^2).
\end{align*}


\item The fact that $\Var(\mathcal{X}_w)\leq \delta_{max}$ follows from direct computation:
\begin{align*}
\Var(\mathcal{X}_w) &= \E[(\mu_0 - \mathcal{X}_w)^2]\\
&= \E(\mathcal{X}_w^2) - \mu_0^2\\
&= \w' \E(\X \X')\w - \w'\one_N \mu_0^2\one_N'\w\\
&= \w' \left[ \E(\X \X') - \mu_0^2 \one_N \one_N' \right]\w\\
&= \w' \E[(\X - \one_N \mu_0)(\X - \one_N \mu_0)']\w\\
&= \w' \Cov(\X)\w\\
&\leq \delta_{max} \one_N'\w\\
&= \delta_{max}.
\end{align*}
To see the identity part of the statement, note that $$\Var(\mathcal{X}_w)  = \w' \Cov(\X)\w = \Sigma_{i=1}^N \Sigma_{j=1}^N w_{ij} \Cov(X_i, X_j),$$ where $w_{ij} = w_iw_j \in [0,1]$ and $\sum_{i=1}^N \sum_{j=1}^N w_{ij} = 1$.  First, suppose that $\Var(X_m) = \delta_{max} > \Var(X_i) = \delta_i$ for all $i \neq m$. Then, if $w_{ii} > 0$ for some $i \neq m$, the term $w_{ii}\Cov(X_i, X_i)$ brings $\Var(\mathcal{X}_w)$ below $\delta_{max}$. This decrease cannot be compensated by any other term because no element in $\Cov(\X)$ is larger than $\delta_{max}$. Consequently, it must be case that $w_i = 0$ for all $i \neq m$. Now, if there exists $j \neq m$ such that $\delta_j = \delta_{max}$ and $w_j > 0$, then $\Var(\mathcal{X}_w) = \delta_{max}$ only if all weight is given to $X_m$ and $X_j$, and $\Cov(X_j, X_m) = \delta_{max}$. This covariance implies that $\Corr(X_j, X_m) = 1$. Thus, $\sigma(X_j) = \sigma(X_m)$ and hence that $X_j = \E[Y | \sigma(X_j)] = \E[Y | \sigma(X_m)] = X_m$. Consequently, $\Var(\mathcal{X}_w) = \delta_{max}$ only if all weight is distributed among $X_i$ such that $X_i = X_m$. 

 From the Theorem \ref{optimal}, $\delta_{max} \leq \Var(\mathcal{X}'')$, where the inequality arises from the Cauchy-Schwarz inequality. It is well-known that this reduces to an equality if and only if $\mathcal{X}''$ and $X_m$ are linearly dependent. Such a linear dependence would imply that $\sigma(\mathcal{X}'') = \sigma(X_m)$ and hence that $X_m = \E[Y | \sigma(X_m)] = \E[Y | \sigma(\mathcal{X}'')] = \mathcal{X}''$. Now, if there exists $j \neq m$ such that $\delta_j = \delta_{max}$, then by the same argument $\sigma(\mathcal{X}'') = \sigma(X_m) = \sigma(X_j)$ and consequently $X_j = X_m = \mathcal{X}''$. 

Putting this all together gives that $\w'\X = \mathcal{X}''$ if and only if $\sigma(X_m) = \sigma(\mathcal{X}'')$ and $w_i > 0$ only for all $X_i = X_m$.

\end{enumerate}
\qed

\subsection{Derivation of Equation \ref{decomposition}}
Suppose that $\mathcal{X}_k \in \{f_1, \dots, f_I\}$ for some finite $I$. Let $K_i$ be the number of times $f_i$ occurs, $\bar{Y}_i$ be the empirical average of $\{Y_k : \mathcal{X}_k = f_i\}$, and $\bar{Y} = \frac{1}{K} \sum_{k=1}^K Y_k$. Then,
\begin{align*}
&\hspace{1.3em}  \frac{1}{K} \sum_{k=1}^K (Y_k - \mathcal{X}_k)^2\\
 &= \frac{1}{K} \left( \sum_{k=1}^K \mathcal{X}_k^2 - 2\sum_{k=1}^KY_k\mathcal{X}_k + \sum_{k=1}^KY_k^2 \right)\\
 &= \frac{1}{K} \left[ \sum_{i=1}^I K_i f_i^2 - 2\sum_{i=1}^I K_i f_i \bar{Y}_i + \left( 2 \sum_{i=1}^IK_i\bar{Y}_i\bar{Y}- 2 \sum_{i=1}^IK_i\bar{Y}_i\bar{Y} \right) \right. \\
 &\qquad \left. {} +  \left( \sum_{i=1}^IK_i\bar{Y}^2- \sum_{i=1}^IK_i\bar{Y}^2 \right) + \sum_{k=1}^KY_k^2 \right]\\
 &= \frac{1}{K} \left[ \sum_{i=1}^I K_i \left( f_i^2 - 2 f_i \bar{Y}_i + 2 \bar{Y}_i\bar{Y}-  \bar{Y}^2 \right) + \sum_{k=1}^K(Y_k^2 -2\bar{Y}_k\bar{Y}+ \bar{Y}^2) \right]\\
 &= \frac{1}{K} \left[ \sum_{i=1}^I K_i \left( f_i^2 - 2 f_i \bar{Y}_i + (\bar{Y}_i^2 - \bar{Y}_i^2) + 2 \bar{Y}_i\bar{Y}-  \bar{Y}^2 \right) + \sum_{k=1}^K(Y_k -\bar{Y})^2 \right]\\
 &= \frac{1}{K} \left[ \sum_{i=1}^I K_i \left( f_i^2 - 2 f_i \bar{Y}_i + \bar{Y}_i^2 \right)  - \sum_{i=1}^I K_i  \left(\bar{Y}_i^2 - 2 \bar{Y}_i\bar{Y} + \bar{Y}^2 \right) + \sum_{k=1}^K(Y_k -\bar{Y})^2 \right]\\
 &=  \frac{1}{K} \sum_{i=1}^I K_i \left( f_i - \bar{Y}_i \right)^2  - \frac{1}{K} \sum_{i=1}^I K_i  \left(\bar{Y}_i - \bar{Y} \right)^2 + \frac{1}{K} \sum_{k=1}^K(Y_k -\bar{Y})^2.
\end{align*}

\bibliographystyle{apalike}
\bibliography{biblio}		

\end{document}